\documentclass[sigconf, anonymous,table,xcdraw]{acmart}
\def\BibTeX{{\rm B\kern-.05em{\sc i\kern-.025em b}\kern-.08emT\kern-.1667em\lower.7ex\hbox{E}\kern-.125emX}}

\usepackage{nicefrac}
\usepackage{siunitx}
\usepackage{array,framed}
\usepackage{booktabs}
\usepackage{
  color,
  float,
  epsfig,
  wrapfig,
  graphics,
  graphicx,
  subcaption
}

\usepackage{setspace}
\usepackage{latexsym,fancyhdr,url}
\usepackage{algpseudocode}
\usepackage{graphics}
\usepackage{xparse} 
\usepackage{xspace}
\usepackage{multirow}
\usepackage{csvsimple}
\usepackage{balance}
\usepackage[utf8]{inputenc} 
\usepackage[T1]{fontenc}    
\usepackage{hyperref}       
\usepackage{url}            
\usepackage{booktabs}       
\usepackage{amsfonts}       
\usepackage{nicefrac}       
\usepackage{microtype}      
\usepackage{lipsum}
\usepackage{graphicx}

\usepackage{makecell}
\usepackage{rotating}
\usepackage[export]{adjustbox}
\usepackage[linesnumbered, ruled, vlined]{algorithm2e}
\usepackage{subcaption}
\usepackage{tikz}
\usepackage{pgfplots}
\usepackage{import}
\usepackage{array}
\usepackage{tabularx}
\usepackage[inline]{enumitem}
\setlist{nolistsep}
\usepackage{float}
\usepackage{stackengine}
\usepackage{multirow}
\usepackage{balance}
\renewcommand{\vec}[1]{\mathbf{#1}}
\newcolumntype{Y}{>{\centering\arraybackslash}X}

\usepackage{hyperref}

\usepackage{textcomp}
\usepackage{amsmath}

\usepackage{amssymb}

\usepackage{mathtools,}

\DeclareMathAlphabet{\mathcal}{OMS}{cmsy}{m}{n}

\DeclareGraphicsExtensions{%
    .png,.PNG,%
    .pdf,.PDF,%
    .jpg,.mps,.jpeg,.jbig2,.jb2,.JPG,.JPEG,.JBIG2,.JB2}

\usepackage{xparse}
\newcommand{\bnm}{\begin{newmath}}
\newcommand{\enm}{\end{newmath}}

\newcommand{\bea}{\begin{eqnarray*}}%
\newcommand{\eea}{\end{eqnarray*}}%

\newcommand{\bne}{\begin{newequation}}
\newcommand{\ene}{\end{newequation}}

\newcommand{\bal}{\begin{newalign}}
\newcommand{\eal}{\end{newalign}}

\newenvironment{newalign}{\begin{align}%
\setlength{\abovedisplayskip}{4pt}%
\setlength{\belowdisplayskip}{4pt}%
\setlength{\abovedisplayshortskip}{6pt}%
\setlength{\belowdisplayshortskip}{6pt} }{\end{align}}

\newenvironment{newmath}{\begin{displaymath}%
\setlength{\abovedisplayskip}{4pt}%
\setlength{\belowdisplayskip}{4pt}%
\setlength{\abovedisplayshortskip}{6pt}%
\setlength{\belowdisplayshortskip}{6pt} }{\end{displaymath}}

\newenvironment{newequation}{\begin{equation}%
\setlength{\abovedisplayskip}{4pt}%
\setlength{\belowdisplayskip}{4pt}%
\setlength{\abovedisplayshortskip}{6pt}%
\setlength{\belowdisplayshortskip}{6pt} }{\end{equation}}

\newcounter{ctr}

\newcounter{mytable}
\def\mytable{\begin{centering}\refstepcounter{mytable}}
\def\endmytable{\end{centering}}

\newcounter{myfig}
\def\myfig{\begin{centering}\refstepcounter{myfig}}
\def\endmyfig{\end{centering}}

\newlength{\saveparindent}
\newlength{\saveparskip}
\newcommand{\E}{{\rm I\kern-.3em E}}

\renewcommand{\eqref}[1]{\mbox{Equation~(\ref{#1})}}

\def \part {part}

\renewcommand{\paragraph}[1]{\vspace*{6pt}\noindent\textbf{#1}\;}

\theoremstyle{definition}
\newtheoremstyle{subdefinition}
  {0pt}       
  {0pt}       
  {\upshape}  
  {0pt}       
  {\itshape} 
  {.}         
  {5pt plus 1pt minus 1pt} 
  {}          
\theoremstyle{subdefinition}
\newtheorem{subdefinition}{Definition}[theorem]

\def \blackslug{\hbox{\hskip 1pt \vrule width 4pt height 8pt
    depth 1.5pt \hskip 1pt}}
\def \qed{\quad\blackslug\lower 8.5pt\null\par}

\newcounter{mynote}[section]

\newcommand\ignore[1]{}

\newcounter{rcnote}[section]

\newcounter{mrnote}[section]

\newcounter{fknote}[section]

\newcounter{anote}[section]

\DeclareMathSymbol{\mlq}{\mathord}{operators}{``}
\DeclareMathSymbol{\mrq}{\mathord}{operators}{`'}

\newcommand{\rhf}[2]{R_{f, \gamma}}

\DeclareDocumentCommand{\edist}{o o}{
  \ensuremath{
    \IfNoValueTF{#1}{{d}}{{\sf d}(#1,#2)}
  }
}

\newcommand{\olrk}[1]{\ifx\nursymbol#1\else\!\!\mskip4.5mu plus 0.5mu\left(\mskip0.5mu plus0.5mu #1\mskip1.5mu plus0.5mu \right)\fi}

\NewDocumentCommand{\indseq}{ O{1} O{r} }{{#1}\ldots {#2}}

\begin{document}
\fancyhead{}
\def\thetitle{\textit{Liuer Mihou}: A Practical Framework for Generating and Evaluating Grey-box Adversarial Attacks against NIDS}
\title{\thetitle}

\author{Ke He}
\affiliation{Faculty of Science\\
    University of Auckland\\
    Auckland, New Zealand \\
    }
\email{khe429@aucklanduni.ac.nz}

\author{Dan Dongseong Kim}
\affiliation{School of ITEE\\
        University of Queensland\\
Brisbane, Australia}
\email{dan.kim@uq.edu.au}

\author{Jing Sun}
\affiliation{Faculty of Science\\
    University of Auckland\\
    Auckland, New Zealand \\
    }
\email{jing.sun@auckland.ac.nz}

\author{Jeong Do Yoo}
\affiliation{School of Cybersecurity\\
Korea University\\
Seoul, Korea}
\email{opteryx25104@korea.ac.kr}

\author{Young Hun Lee}
\affiliation{School of Cybersecurity\\
Korea University\\
Seoul, Korea}
\email{dlddudfkr@korea.ac.kr}

\author{Huy Kang Kim}
\affiliation{School of Cybersecurity\\
Korea University\\
Seoul, Korea}
\email{cenda@korea.ac.kr}

\date{}

\begin{abstract}
Due to its high expressiveness and speed, Deep Learning (DL) has become an increasingly popular choice as the detection algorithm for Network-based Intrusion Detection Systems (NIDSes). 
Unfortunately, DL algorithms are vulnerable to adversarial examples that inject imperceptible modifications to the input and causes the DL algorithm to misclassify the input. 
Existing adversarial attacks in the NIDS domain often manipulate the traffic features directly, which hold no practical significance because traffic features cannot be replayed in a real network.
It remains a research challenge to generate practical and evasive adversarial attacks.

This paper presents the \textit{Liuer Mihou} attack that generates practical and replayable adversarial network packets that can bypass anomaly-based NIDS deployed in the Internet of Things (IoT) networks. 
The core idea behind \textit{Liuer Mihou} is to exploit adversarial transferability and generate adversarial packets on a surrogate NIDS constrained by predefined mutation operations to ensure practicality. 
We objectively analyse the evasiveness of \textit{Liuer Mihou} against four ML-based algorithms (LOF, OCSVM, RRCF, and SOM) and the state-of-the-art NIDS, Kitsune. 
From the results of our experiment, we gain valuable insights into necessary conditions on the adversarial transferability of anomaly detection algorithms. 
Going beyond a theoretical setting, we replay the adversarial attack in a real IoT testbed to examine the practicality of \textit{Liuer Mihou}.  
Furthermore, we demonstrate that existing feature-level adversarial defence cannot defend against \textit{Liuer Mihou} and constructively criticise the limitations of feature-level adversarial defences.
\end{abstract}

\ccsdesc[500]{Security and privacy~Network security}
\ccsdesc[500]{Security and privacy~Intrusion detection systems}
\ccsdesc[300]{Computing methodologies~Artificial intelligence}

\keywords{NIDS, Deep Learning, Adversarial Attacks}

\maketitle


\section{Introduction}
Deep Learning (DL) has gained notable popularity over recent years as it outperforms traditional Machine Learning (ML) algorithms across various domains such as Natural Language Processing (NLP) \cite{sak2014long} and Computer Vision (CV) \cite{bochkovskiy2020yolov4, he2016identity}. 
The success of DL is primarily due to its ability to utilise large volumes of data to learn highly abstract representations that are incomprehensible to traditional ML algorithms and humans\cite{lecun2015deep}, making them a perfect fit for Network-based Intrusion Detection Systems (NIDSes) where large volumes of data are generated each day. 

Unfortunately, a significant weakness of DL is that it is vulnerable to adversarial examples.
Adversarial examples are created by purposely adding an imperceptible perturbation to an input of a DL algorithm that causes misclassification of the input by the DL algorithm. 
Various adversarial attacks against DL algorithms have been developed across multiple domains such as images \cite{szegedy2013intriguing, papernot2016limitations}, physical world \cite{kurakin2016adversarial}, and malware \cite{hu2017generating}. 

The use of Deep Neural Networks (DNNs) in NIDS exposes a new attack surface for attackers to exploit and is particularly devastating because they are under constant adversarial threats \cite{sommer2010outside}. 
Most adversarial attacks against NIDS have focused on modifying the network features directly \cite{lin2018idsgan, clements2019rallying, ibitoye2019analyzing, piplai2020nattack}. 
However, feature-level adversarial attacks are not practical since they only modify the traffic features, which cannot be replayed directly in the network to conduct the intended malicious activities. 
Hence, practical adversarial attacks against NIDS should consider \textit{problem/packet space} \cite{pierazzi2020intriguing} modifications that directly alter the packets.

Previous works have made limited progress towards practical packet-level adversarial attacks. 
Early packet-level attacks often rely on randomly applying predefined mutations with vigorous trial-and-error to find a feasible solution \cite{homoliak2018improving, hashemi2019towards}, which provide little to no theoretical guidance and insight. 
Recent works in packet-level adversarial attacks \cite{han2020practical, kuppa2019black} formulate the adversarial attack as a bi-level optimisation problem that first searches for the adversarial features and then modifies the packets to mimic the adversarial features. 
However, the bi-level design is overly complicated and introduces numerical instabilities that make finding the optimal solution challenging.

In this paper, we present a novel and practical transfer-based \cite{papernot2016transferability} adversarial attack called \textit{Liuer Mihou}\footnote{\textit{Liuer Mihou}, also known as the six-eared macaque, is one of the antagonists in the classic Chinese novel \emph{Journey to the West} and is best known for imitating \textit{Sun Wukong}, the monkey king.} to attack NIDS. 
\textit{Liuer Mihou} first trains a surrogate NIDS that mimics the decision boundaries of the target NIDS. 
Next, the attack finds the optimal set of semantic preserving packet mutations with a hybrid heuristic search algorithm to minimise the anomaly score of any packets identified as malicious by the surrogate NIDS. 
As a result, \textit{Liuer Mihou} produces a realistic sequence of packets that can be replayed in real-time in the network.

To fully understand the strength and weaknesses of \textit{Liuer Mihou}, we comprehensively evaluate the evasiveness of \textit{Liuer Mihou} in a real Internet of Things (IoT) network against a wide range of ML/DL based NIDS.
Going beyond the theoretical setting, we replayed the generated adversarial traffic in the original IoT network to evaluate its evasiveness and maliciousness.
Finally, we examine the strength of our attack by launching our attack against adversarial defences such as feature squeezing \cite{xu2017feature} and Mag-Net \cite{meng2017magnet}.

\textbf{Contributions.} 
In summary, our key contributions include:
\begin{enumerate}
    \item We design a novel and practical adversarial attack tailored to attack anomaly-based NIDS, called \textit{Liuer Mihou}, and provide source code for download \cite{lm_github}.

    \item We conduct a comprehensive evaluation of \textit{Liuer Mihou} in a real IoT testbed against four ML based anomaly detection algorithms (SOM, RRCF, LOF, and OCSVM), and the state-of-the-art DL based IoT NIDS, Kitsune \cite{mirsky2018kitsune}.
    
    \item We demonstrate the strength of \textit{Liuer Mihou} by assessing our attacks on Kitsune with adversarial detection defences such as Feature Squeezing \cite{xu2017feature} and Mag-Net \cite{meng2017magnet}. 
    
    \item We provide empirical results and findings of our experiments, which provides insights for future adversarial attacks against NIDS.
\end{enumerate}

\textbf{Organisation.} 
The rest of this paper is organised as follows. 
Section \ref{sec:proposed_solution} presents the \textit{Liuer Mihou} framework along with definitions and the threat model used.
Section \ref{sec:eval} describes the experiment setup up, followed by the experiment results presented in Section \ref{sec:exp_res}. 
Next, Section \ref{sec:background} provides background and related work on anomaly-based NIDS, adversarial attacks and defences in the NIDS domain. 
Then, we discuss our findings, limitations, and future work in Section \ref{sec:discussion}. 
Finally, we conclude the paper in Section \ref{sec:conclusion}. 
We have also provided additional information in the Appendix to provide more details of our experiments.
\begin{figure*}[t]
    \centering
    \includegraphics[width=\linewidth]{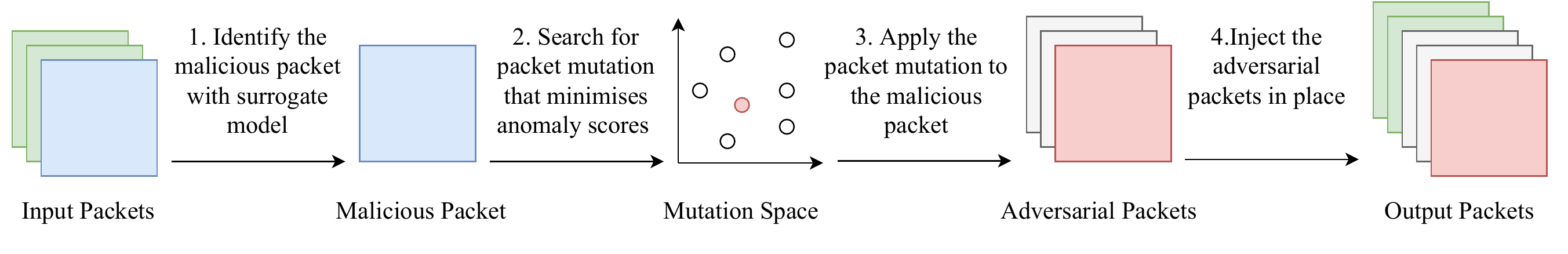}
    \caption{Overview of the \textit{Liuer Mihou} attack. 
    The surrogate model classifies each packet as either benign (green) or malicious (blue). 
    For each malicious packet, \textit{Liuer Mihou} finds an optimal mutation operation that minimises the anomaly scores produced by the surrogate model. 
    Next, the mutation operation is applied to the malicious packet to generate a set of adversarial packets, including the original packet (red) and redundant packets (grey). 
    Finally, the benign and adversarial packets are written in place to the output file.}
    \label{fig:framework_overview}
\end{figure*}

\section{Liuer Mihou}
\label{sec:proposed_solution}
This section first provides definitions of the terms used throughout the paper and explicitly defines the threat model of \textit{Liuer Mihou}. Following that, we present an overview of \textit{Liuer Mihou} and provide details of the critical components.

\subsection{Definitions}
\begin{definition}[Network Traffic Space]
We refer to the Network Traffic Space as a set that contains all the possible packets that NIDS can capture, denoted as $\mathcal{P}$. 
Depending on the nature of the network traffic, $\mathcal{P}$ can be further classified into five categories:
\textit{benign, malicious, clean, adversarial}, and \textit{replay}.

\begin{subdefinition}[Benign Traffic Space ($\mathcal{P}_b$)]
$\mathcal{P}_b$ contains all packets captured during the normal operational time.
\end{subdefinition}

\begin{subdefinition}[Malicious Traffic Space ($\mathcal{P}_m$)]
\label{def:mal_traffic}
$\mathcal{P}_m$ contains all packets captured when the network is under attack. 
Note that not all packets in malicious traffic space are directly responsible for the malicious activities and may contain packets in common with benign traffic (\textit{e.g.,} packets generated with TCP three-way handshake).
\end{subdefinition}

\begin{subdefinition}[Clean Traffic Space ($\mathcal{P}_c$)]
$\mathcal{P}_c$ contains all packets that has not been adversarially modified, \textit{i.e.} all of the benign and malicious traffic ($\mathcal{P}_c=\mathcal{P}_b \cup \mathcal{P}_m$).
\end{subdefinition}

\begin{subdefinition}[Adversarial Traffic Space ($\mathcal{P}_a$)]
$\mathcal{P}_a$ contains all theoretically crafted packets that are generated by applying mutation operations on malicious traffic. 
\end{subdefinition}

\begin{subdefinition}[Replay Traffic Space ($\mathcal{P}_r$)]
$\mathcal{P}_r$ contains all packets captured during the replay of the adversarial traffic. 
Note that we make an distinction between $\mathcal{P}_r$ and $\mathcal{P}_a$ because inherent transmission and processing delays during replay causes the traffic pattern for $\mathcal{P}_r$ to be different compared to $\mathcal{P}_a$, shown later in Figure \ref{fig:kitsune_as}.
\end{subdefinition}
\end{definition}

\begin{definition}[Feature Extraction]
Feature extraction ($\varphi : \mathcal{P} \longrightarrow \mathcal{X} \in \mathbb{R}^n$) is a function that extracts $n$ dimensional features $ x \in \mathcal{X}$ from a network packet $p$ such that $\varphi(p)=x$.
Due to the sheer size of network traffic, NIDS often extracts aggregate information from the network traffic and uses the network features instead of network packets for detection.
\end{definition}

\begin{definition}[NIDS]
The NIDS ($g : \mathcal{X} \longrightarrow \mathcal{A} \in \mathbb{R}$) can be modelled as a function that produces a one-dimensional anomaly score ($a \in \mathcal{A}$) based on the input features ($x$). 
If the anomaly score is above a predefined threshold ($t$), $g$ classifies the corresponding $p_i$ as malicious, otherwise $p_i$ is benign.
\end{definition}

\begin{definition}[Mutation Operations]
Mutation operations ($ \mathcal{M} : \mathcal{P}_c \longrightarrow \mathcal{P}_a$) take a clean packet $p \in \mathcal{P}_c$ and modify it to adversarial packets ${P^\prime \in \mathcal{P}_a}$. 
Note that the adversarial packets $P^\prime$ is a set because the mutation operations may inject several redundant packets before $p$.
\end{definition}

\subsection{Threat Model}
\label{sec:threat_model}
We strictly target outlier detection based NIDS instead of classification based NIDS because they are more practical. 
The abundance of network traffic generated every day makes labelling time-consuming, and correctly labelling the traffic requires expert knowledge.   
Furthermore, we only consider packet-level features because flow-level features need to wait for the connection to finish before producing the features, which is not practical for real-time detection.  

The attacker's goal, knowledge, and capability are defined in the following and are in line with prior works in adversarial attacks against NIDS \cite{han2020practical, lin2018idsgan, homoliak2018improving}: 
\begin{description}
\item [Goal] 
The attacker's goals are two-fold: the attacker wishes to fully/partially maintain the security violations caused by the malicious attack and have the malicious attack being classified as benign by the target NIDS. 

\item [Knowledge] 
The attacker operates under a grey-box setting with complete knowledge of the network features extracted ($\mathcal{X}$) but knows nothing about the classifier ($g$), except that it uses an outlier detection algorithm. 
Knowing the features extracted is a reasonable assumption, as feature extractors used in NIDS often extract similar features, such as statistics of arrival time and payload size over various time intervals \cite{han2020practical}. 
Moreover, feature extractors for most recent datasets are publicly available online, \textit{e.g.,} CIC-FlowMeter \cite{sharafaldin2018toward} and AfterImage \cite{mirsky2018kitsune}. 

\item [Capability] 
We assume the attacker is inside the network and can sniff both benign and malicious traffic. 
In IoT networks where devices communicate wirelessly with little to no encryption, packets can be sniffed easily with a wireless sniffer.
The attacker can also modify and inject crafted packets in malicious traffic and replay the modified traffic in the network.
\end{description}

Under our threat model, the attacker can efficiently train a surrogate NIDS, $g^\prime$, with a threshold, $t^\prime$, with arbitrary architecture based on benign traffic ($P_b$).

\subsection{Overview of Liuer Mihou}
\textit{Liuer Mihou} is a practical adversarial generation algorithm tailored specifically for NIDS.
It operates iteratively on each packet of the malicious traffic, illustrated in Figure \ref{fig:framework_overview}.
For each packet, the surrogate NIDS first classifies the packet. 
Suppose the packet is classified as benign by the surrogate. In that case, the packet is likely to be classified as benign by the target NIDS, so no modifications are needed, and \textit{Liuer Mihou} writes the packet to the output straight away (green). 
On the other hand, if the surrogate classifies the packet as malicious, it is likely to be classified as malicious by the target model (blue). 
For each malicious packet identified by the surrogate, \textit{Liuer Mihou} first searches for an optimal set of mutation operations on the packet that minimises the anomaly score produced by the surrogate NIDS (which also reduces the anomaly scores produced by the target NIDS). 
Next, the optimal mutation operations are applied to the malicious packet, transforming it into adversarial packets containing the modified malicious packet (red) and some redundant packets (grey) as byproducts. 
Finally, the adversarial packets are written to the output file in place. 

\subsection{Mutation Operations}
\label{sec:mutation_operations}
In order to manipulate the packets, we have to define a set of mutation operations, $\mathcal{M}$ that can be applied to a malicious packet to change the extracted features of the packet with minimal change of the content.
The choice of $\mathcal{M}$ will depend mainly on the feature extractor, and by inspecting existing open-source traffic feature extractors \cite{mirsky2018kitsune, sharafaldin2018toward}, we have found that most feature extractors measure statistics of the inter-arrival time and packet size. 
Therefore, we propose two simple mutation operations that change these two features:
\begin{description}
\item [Packet Delay] Delay the arrival time of a packet, which changes the inter-arrival time distribution. 
\item [Packet Injection] Inject redundant packets before a packet in the same connection, which changes packet-size distribution.
\end{description}

The maximum time delay and the maximum number of redundant packets injected are constrained to ensure the adversarial traffic does not differ too much from the malicious traffic and reduces the search space's size. Full set of \textit{Liuer Mihou} hyperparameters is presented in Appendix \ref{sec:hyperparameters}.

Pierazzi \textit{et al.} \cite{pierazzi2020intriguing} proposed four general constraints for problem space modification, and we show our mutation operations satisfies these constraints.
\begin{enumerate}
    \item \textit{Available Transformations.}
    It is trivial to see that an attacker can easily delay and inject redundant packets under our threat model.
    
    \item \textit{Preserved Semantics.} 
    Our mutation operations do not modify the malicious packets' payload, preserving the original intended malicious activity. For attacks that rely on the packets' inter-arrival time, such as DoS attacks, we place constraints on the mutation operations to adjust the maliciousness of the adversarial attack.
    
    \item \textit{Plausibility.} 
    We have ensured the plausibility of the mutation operation by manually checking the adversarial packets in Wireshark and replaying the packet in the same network, and there are no packets that seem blatantly abnormal.
    
    \item \textit{Robustness to Preprocessing.}
    A common non-ML based preprocessing in NIDS is to block the attacker's IP by the victim. Under such circumstances, the attacker can spoof its IP address to bypass blocking.
\end{enumerate}

\subsection{Attack Objective Function}
The objective function of \textit{Liuer Mihou} is formulated as the following optimisation problem, done for each malicious packet ($p_i$):
\begin{equation}
\label{opt_max}
\min_{m\in\mathcal{M}} \max \{ g^\prime(\varphi(p^\prime)) : p^\prime \in m(p_i) \}
\end{equation}
where $\mathcal{M}$ is the space of all possible mutation operations, $m$ is a specific mutation operation, $p_i$ is the $i^{th}$ malicious packet, $p^\prime$ is the set of adversarial packets after mutation. 
For simplicity and brevity, we refer to the value of $\max \{ g^\prime(\varphi(p^\prime)) : p^\prime \in m(p_i) \}$ as the cost value, denoted by $c$.

Intuitively, Equation \eqref{opt_max} aims to directly minimise the maximum anomaly score of the adversarial packets that are obtainable with mutation operations on the malicious packet. 
Since \textit{Liuer Mihou} operates under the grey-box scenario, we calculate the anomaly scores based on a surrogate NIDS ($g^\prime$) that uses an arbitrary outlier detection algorithm. 
Although the surrogate NIDS will have a slightly different decision function and anomaly threshold, we still expect adversarial examples that bypass the surrogate NIDS to also bypass the target NIDS due to adversarial transferability \cite{papernot2016transferability}.

Notice that the adversarial packets contain the modified malicious packet and several redundant packets. We wish to reduce the maximum anomaly score of the adversarial packets so that we are not introducing more malicious packets.  
Moreover, we only minimise the cost value in our attack formulation without explicitly placing the constraint $c<t^\prime$ to allow a more flexible generation of the adversarial traffic. 
Consider a scenario where the search algorithm failed to find a mutation operation with the cost value above the surrogate threshold due to tight boundary constraints or limited search time. 
If we have introduced the constraint $c<t^\prime$, we will have no feasible solution. 
Thus, we only search for the mutation operation with the lowest anomaly scores to make the attack more flexible. 
In the case where there are still malicious packets above the surrogate threshold, we can recursively run \textit{Liuer Mihou} on the output adversarial packets until \textit{Liuer Mihou} reduces all packets below the surrogate threshold.

\subsection{Packet Vectorisation}
\label{sec:vec}
To efficiently search for the mutation operations, we abstract the representation of mutation operations into low-dimensional vectors in the mutation space ($\Psi$). 
Each vector in the mutation space represents a set of mutation operations applied to the malicious packet, and optimisation of Equation \eqref{opt_max} is done by moving the vectors in the mutation space. 

With the mutation operations defined in Section \ref{sec:mutation_operations}, we define $\Psi$ as a two-dimensional space with ($t_m, n_c$):
\begin{itemize}

\item Modified arrival time of the packet ($t_m$), which represents \textbf{packet delay}.

\item Number of redundant packets inserted before the packet ($n_c$), which represents \textbf{packet injection}.

\end{itemize}
For example, suppose the optimal vector is $(0.4,4)$ in the mutation space. In that case, \textit{Liuer Mihou} will delay the malicious packet by 0.4 seconds and place four redundant packets before the malicious packet.

Applying packet delay can be trivially achieved by changing the packet's arrival time. 
However, packet injection is more complicated because we must define each redundant packet's arrival time and payload size. 
We have experimented with the following three methods of assigning arrival time and payload size of the redundant packets.

\begin{figure}[t]
    \begin{subfigure}[t]{0.32\linewidth}
        \includegraphics[width=\linewidth]{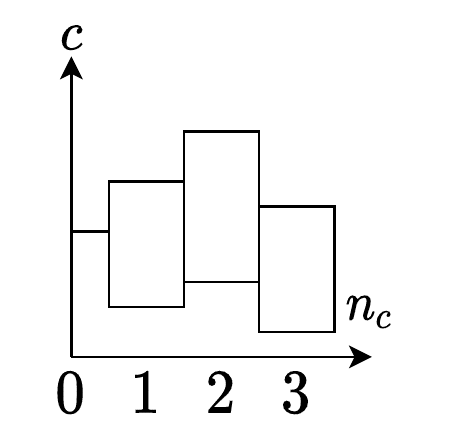}
    \caption{RA}
    \label{fig:2d_vec}
    \end{subfigure}
    \hfill
    \begin{subfigure}[t]{0.32\linewidth}
        \includegraphics[width=\textwidth]{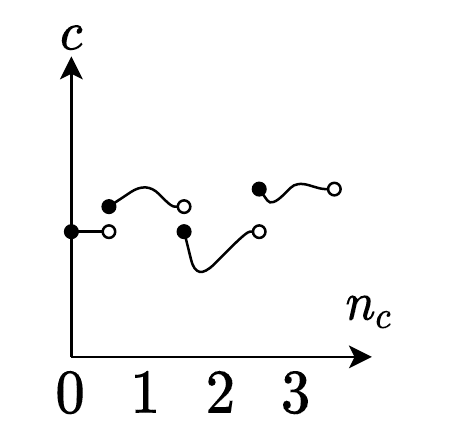}
    \caption{SA}
    \label{fig:s2d_vec}
    \end{subfigure}
    \hfill
    \begin{subfigure}[t]{0.32\linewidth}
        \includegraphics[width=\linewidth]{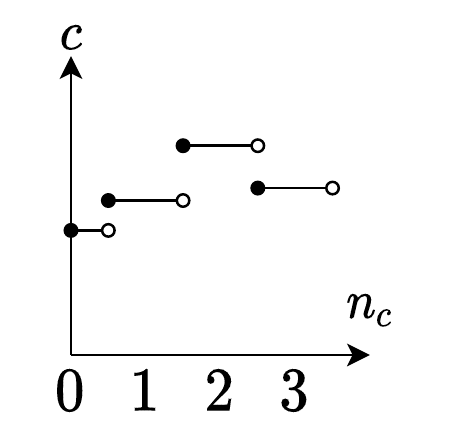}
    \caption{UA}
    \label{fig:3d_vec}
    \end{subfigure}
    \caption{Illustration of cost values for three assignment methods, Random (RA), Seeded (SA) and Uniform (UA). 
    The y-axis represents the cost function, $c$, and the x-axis represents $n_c$. 
    All other variables (\textit{i.e.,} $t_m, s_c$) are held constant.}
    \label{fig:vec_vis}
\end{figure}

\textbf{Random Assignment (RA).} 
RA randomly assigns the arrival time and payload size of each redundant packet.
We have found this method causes the cost function to be non-deterministic because the arrival time and payload size of the redundant packets at the same position are different in each iteration, as shown in Figure \ref{fig:2d_vec}.
The non-deterministic nature of the cost function makes it difficult for the search algorithm to find the optimal solution.

\textbf{Seeded Assignment (SA).}
SA seeds the random number generator with the value of $n_c$ before generating the payload sizes for each redundant packet. 
The goal of seeding is so that the cost function for any real value of $n_c$ will be deterministic and continuous in the interval between $[n_c-0.5, n_c+0.5)$, shown in Figure \ref{fig:s2d_vec}. 
Intuitively, this allows flexible assignment of packet sizes and payload while making the cost function deterministic.
The arrival time of the redundant packets is assumed to be evenly spaced between the arrival time of the previous packet and $t_m$. 

\textbf{Uniform Assignment (UA).} 
A new dimension is introduced in the mutation space, $s_c$, which governs the payload size of all redundant packets. 
As a result, the cost value is deterministic and piece-wise constant around each integer value of $n_c$ (shown in Figure \ref{fig:3d_vec}). 
Similar to the seeded assignment, the arrival time of crafted packets is uniformly distributed. 

We have experimented with assigning redundant packets with random arrival time, resulting in worse performance. 
We hypothesise that consistent inter-arrival time of packets is characteristic of benign traffic in our dataset. 
Therefore, redundant packets with consistent inter-arrival time will always produce lower costs. 
Similarly, our auxiliary experiments in Appendix \ref{sec:finding_opt_comb} showed UA is the best method for creating redundant packets, which suggest that uniform payload sizes are also characteristic of benign traffic in our dataset. 

\begin{algorithm}[t]
 \KwData{A packet $p_i$}
 \KwResult{Mutation operation ($m \in \mathcal{M}$) that generate minimum cost $c$ in Equation \eqref{opt_max}}
 Define the bounded mutation space, $\Psi$\;
 Randomly initialise the particle position $\vec{x}$ and velocity $\vec{v}$ of the population within $\Psi$\;
 \For{$i \gets 0 $ \KwTo max\_iter}{
  Calculate the cost of each particle $c$\;
  \eIf{random.uniform() $<$ mutation probability}{
   \ForEach{particle $x_i$ in population}{
  Identify the best position\;
  Find the neighbourhood best position\;
  Calculate the velocity by adding inertia weight, cognitive term, and social term\;
  Update position with velocity\;
  }
   }
  {
     \ForEach{particle $x_i$ in population}{
   Calculate the mutant $m$\;
   Calculate the candidate particle $x_m$\;
   Calculate the cost of $x_m$ \;
   Replace $x_i$ with $x_m$ if $x_m$ has a lower cost\;
   }
  }
 }
 Transform the best particle position back to $\mathcal{M}$ and return the mutation operation\;
 \caption{The PSO-DE algorithm.}
  \label{alg:pso_de}
\end{algorithm}

\subsection{Search Algorithms}
\label{sec:search}
The standard methods for optimising feature-level attack formulation are gradient descent algorithms \cite{szegedy2013intriguing, papernot2016limitations, moosavi2016deepfool,madry2017towards}.
However, gradient descent algorithms cannot be applied to Equation \eqref{opt_max} because $\varphi$ is non-differentiable and non-invertible, commonly known as the inverse feature mapping problem \cite{pierazzi2020intriguing}. 
In addition, our mutation operations may inject redundant packets and further complicate gradient descent. 
The typical approach to solve optimisation problems involving non-differentiable and non-invertible functions is to use meta-heuristic algorithms \cite{blum2003metaheuristics}. 
Meta-heuristic algorithms often have a master strategy that iteratively generates solutions, and the optimal solution is discovered by continuously evolving the solutions according to the fitness function. 
All meta-heuristic algorithms have an inevitable trade-off between exploration and exploitation, and finding a balance between the trade-off will usually ensure global optimality is achievable. 
Some of the well-known meta-heuristic algorithms include Differential Evolution (DE) \cite{storn1997differential}, Ant Colony Optimisation (ACO) \cite{dorigo2006ant}, PSO \cite{kennedy1995particle}, and Grey Wolf Optimiser (GWO) \cite{mirjalili2014grey}.

\textit{Liuer Mihou} utilises a hybrid heuristic search algorithm that combines PSO and DE to search for the optimal mutation, which we refer to as PSO-DE. 
Our auxiliary experiments in Appendix \ref{sec:opt_with_pso} show vanilla PSO is particularly prone to get stuck in local optima, known as the stagnating particles problem \cite{engelbrecht2013particle}.
Hence, DE is introduced to increase the exploration ability of the search algorithm.
The pseudocode for PSO-DE is presented in Algorithm \ref{alg:pso_de}. 
The mutation vectors are called particles, which is conventional in heuristic optimisation literature.
The initial position of the particles are generated randomly within the mutation space (lines 1 to 2). 
In each iteration, each particle's cost is calculated with Equation \eqref{opt_max}, and the particle position is stochastically updated using either PSO or DE (lines 3 to 5), governed by mutation probability. 

PSO updates its particles by moving each particle according to their velocity (lines 6 to 10). The velocity is calculated additively with three components:
\begin{description}

    \item[Inertia weight] Percentage of the original velocity kept unchanged, calculated with $w \times v_c$.
    
    \item[Cognitive term] Velocity towards the best solution, calculated with $c_1 r_1 (x_{pbest}-x)$ (exploitation).
    
    \item[Social term] Velocity towards the neighbourhood best solution, calculated with $c_2  r_2  (x_{best}-x)$ (exploration).
    
\end{description}
where $w,c_1,c_2$ are hyperparameters, $v_c$ is the current velocity, $r_1, r_2$ are random values between 0 to 1, and $x_{pbest}, x_{best}, x$ are position of personal best, neighbourhood best, and current position, respectively. 

DE updates its particles via a sequence of mutation, crossover and update operations (lines 12 to 16):
\begin{description}

    \item[Mutation] Choose three other particles, $x_a, x_b, x_c$ and calculate the mutant $m$ with mutation factor $\alpha$, $m=x_a+\alpha(x_b-x_c)$.
    
    \item[Crossover] Calculate the candidate particle by combining mutant with original particle. 
    Each dimension has probability $p_r$ to have the mutant value and $1-p_r$ to have the original particle. 
    
    \item[Update] If the original particle has a higher cost, it is replaced by the candidate particle.
\end{description}

We have conducted comprehensive experiments on finding optimal combination of search algorithm and payload assignment strategies and have found UA with PSO-DE performs best overall.
Details of the experiments are provided in Appendix \ref{sec:finding_opt_comb}.

\section{Experimental Setup}
\label{sec:eval}
This section describes the experimental setup we used to evaluate \textit{Liuer Mihou}.
We begin with a description of the dataset, followed by the details of the target ML/DL algorithms and metrics used.
\begin{figure}[t]
    \centering
    \includegraphics[width=\linewidth]{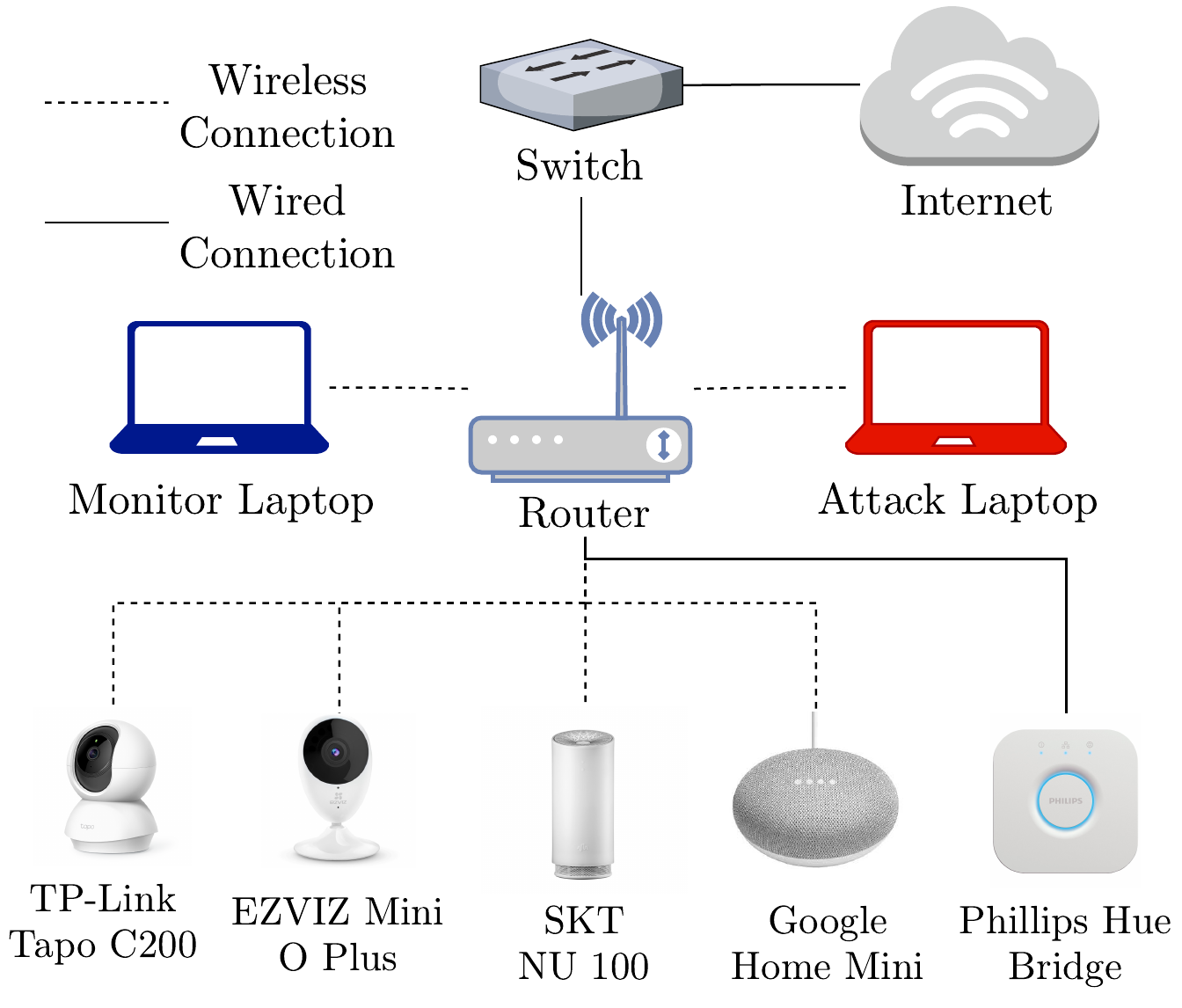}
    \caption{The IoT testbed used in our experiments.}
    \label{fig:iot_testbed}
\end{figure}

\subsection{Dataset}
We evaluate \textit{Liuer Mihou} against IoT networks.
IoT is one of the fastest-growing technologies in the history of computing and is becoming prevalent in smart cities and smart homes \cite{al2020survey}. 
In addition, IoT devices have low computation power and transmit unencrypted traffic, making them highly vulnerable to attacks. 
Therefore, it is of great interest to both adversarially attack and defend IoT networks.

We use two IoT networks in our experiments, one being the benchmark Kitsune dataset\cite{mirsky2018kitsune} and another is a dataset captured from our own IoT testbed \cite{kudataset}.
Measuring the maliciousness of the adversarial traffic is a crucial part of our evaluation.  
However, replaying the adversarial attack generated with any publicly available network datasets is impossible because we cannot fully replicate the testbed used to capture the data. 
Hence, we have gathered our dataset to evaluate the maliciousness of the adversarial attacks.
The testbed consists of various IoT devices connected wirelessly, including AI Speakers (Google Home, NUGU), Security Cameras (EZVIZ, Tapo), and a Smart Hub (Hue Bridge). 
The testbed also includes an attacker laptop used to conduct attacks and a monitor laptop used to capture network packets. 
Each device can be controlled or monitored by a cell phone, PC, or other connected IoT devices.
Figure \ref{fig:iot_testbed} illustrates the setup of the IoT testbed.

The benign data consists of packets generated via commands such as turning off light bulbs, playing music on Google Home, \textit{etc}, over 30 minutes. 
The malicious data consists of two main types of malicious attacks, probing and Denial of Service (DoS), all targeted at Google Home Mini. 
Distributed attacks such as Mirai Botnet were excluded since it requires compromising other IoT devices, which is outside the scope of our threat model.
Probing attacks include Port Scan (PS) and OS Detection (OD), and we repeat PS twice and OD four times. 
DoS attacks include HTTP Flooding (HF) with LOIC at the highest intensity for 6 minutes. 
The attacks were chosen because they are frequently conducted and can be successfully detected by anomaly-based NIDS\footnote{Other common attacks such as ARP poisoning and ARP host discovery were also conducted, but our both victim and surrogate NIDS failed to detect such attacks and is ignored in our experiments}. 
To reduce the training time, we have filtered the packets so that \textit{Liuer Mihou} only processes packets with Google Home as sender or receiver, which results in 14,400 benign packets. 

The adversarial traffic is generated with \textit{Liuer Mihou} with the default, rule-of-thumb parameters that show good performance in general, shown in Appendix \ref{sec:hyperparameters}. 
We intentionally chose not to fine-tune any of the hyperparameters to suit our dataset in order to remove any selective data snooping \cite{arp2020and}.
To reduce the computational time of generating adversarial HTTP Flooding traffic, we only process approximately the first 6\% of the original HTTP Flooding traffic, which generates around 36,000 packets. 
To evaluate the maliciousness of the adversarial traffic, we have replayed the adversarial packets in the same IoT testbed with tcpreplay \cite{tcpreplay} in real-time and captured the replayed packets. 
The number of malicious, adversarial, and replayed packets gathered for each attack can be found in Table \ref{tab:packet_num}.

\subsection{Target DL/ML Algorithms}
\label{sec:exp_setup}
We evaluate our attack on a wide range of DL/ML based anomaly detectors.
In particular, the following anomaly detection algorithms are evaluated.
\begin{itemize}
    \item Kitsune \cite{mirsky2018kitsune}, a state-of-the-art NIDS for IoT network, implemented with Github implementation \cite{kitsune_git}.
    
    \item Self-Organising Maps (SOM) \cite{kohonen1998visual} implemented with the python package MiniSOM \cite{vettigliminisom}.

    \item Robust Random Cut Forest (RRCF) \cite{guha2016robust} implemented with the python package rrcf \cite{bartos_2019_rrcf}.
    
    \item Local Outlier Factor (LOF) implemented with sklearn.

    \item One-Class SVM (OCSVM) implemented with sklearn.
    
    \item Isolation Forest and Elliptical Envelope implemented with sklearn. 
    However, these two algorithms did not perform well on our dataset, see details in Appendix \ref{sec:pre_transfer}. 
\end{itemize}

All algorithms were trained on entire benign samples and used the default parameters. 
Where applicable, the upper bound on the fraction of training errors is set to 0.001, \textit{i.e.}, the operating point of FPR is 0.001 or less.

The surrogate NIDS is a vanilla autoencoder written in TensorFlow 2 \cite{tf2anomaly}. 
The encoder consists of three dense layers with 32, 8, and 2 neurons, respectively, and the Decoder consists of three dense layers with 8, 32, and 100 neurons. 
All layers use the ReLU activation function except for the last layer of Decoder, where it uses Sigmoid.
The architecture of the surrogate model is arbitrarily chosen, and we intentionally did not conduct any hyperparameter search to find the optimal structure.
The surrogate model is trained on all benign packets with one epoch to mimic online detection. 
The threshold of the surrogate model is determined as three standard deviations away from the mean of the anomaly scores on benign data. 

The detection results of all traffic are generated offline. This is due to the open-source implementation of the feature extractor, AfterImage \cite{kitsune_git}, is not optimised and is too slow for real-time feature extraction.
Nonetheless, the detection speed of the NIDS is outside the scope of our work, and the accuracy will be the same regardless of online or offline detection.

\subsection{Metrics}
Prior studies in adversarial attacks mainly target classifiers and measure the precision and recall of the model with ground truth labels. 
However, since our attack targets anomaly detectors with no ground truth labels, we cannot use precision and recall as metrics. 
Instead, we propose three categories of metrics to evaluate various aspects of the NIDS under the influence of \textit{Liuer Mihou}: performance, evasion, and semantic. 
A summary of metrics and symbols is provided in Appendix \ref{sec:metrics_symbols}.

\subsubsection{Performance Metrics}
Performance metrics measure the accuracy of NIDS under clean traffic. 
Following the conventions in security literature, benign examples are referred to as negatives and malicious examples as positives.
Performance metrics include True Negative Rate (TNR), which measures the ratio of packets in benign traffic correctly identified as benign packets, and Malicious Detection Rate (MDR), which measures the ratio of packets classified as malicious in the malicious traffic. 
Note that by Definition \ref{def:mal_traffic}, not all packets in malicious traffic are malicious and may contain benign packets. 
Since we do not have the ground truth labels for each malicious packet, we use MDR instead of True Positive Rate. 
A high MDR and TNR indicate that the NIDS can effectively separate malicious and benign traffic.

\subsubsection{Evasion Metrics}
Evasion metrics measure the accuracy of the NIDS under adversarial and replay traffic and consist of two metrics, Detection Rate (DR) and Evasion Rate (ER), with each metric measured for adversarial traffic (ADR and AER) and replayed traffic (RDR and RER). 
DR measures the ratio of the adversarial/replayed packets that have been classified as malicious and gives an indication of the robustness of NIDS under adversarial/replayed traffic. 
However, using DR alone can sometimes be misleading. Consider an attack with an MDR of 0.1, and after adversarial modification, its ADR is reduced to 0.09. 
Judging by ADR alone, we might conclude that the perturbation is working very well, but since the MDR is 0.1, our attack has only reduced 10\% of the initially detected packets. 
Hence, we require ER that measures the percentage of the adversarial/replayed packets that evade detection compared to the original attack, indicating the effectiveness of the adversarial attack. 

\subsubsection{Semantic Metrics}
\label{sec:semantic_metrics}
Semantic metrics compare the severity of the adversarial traffic on the target system to the original, unmodified attack. 
Different network attacks have different goals so the semantic metrics will depend largely on the network attack. 
For DoS attacks, we measure the Round-Trip Time (RTT) and calculate the Relative Round-trip Delay (RRD) of the device under flooding attacks compared to the normal environments. 
For Probing attacks, we compare the Relative Ports Scanned (RPS) and Relative Time Delay (RTD) between the adversarial and unmodified attacks.

\section{Experiment Results}
\label{sec:exp_res}
This section presents the results and findings of the following three main aspects of \textit{Liuer Mihou}: \textbf{evasiveness, maliciousness}, and \textbf{evasiveness against adversarial defences}. 

\begin{figure}[t]
    \centering
    \includegraphics[width=\linewidth]{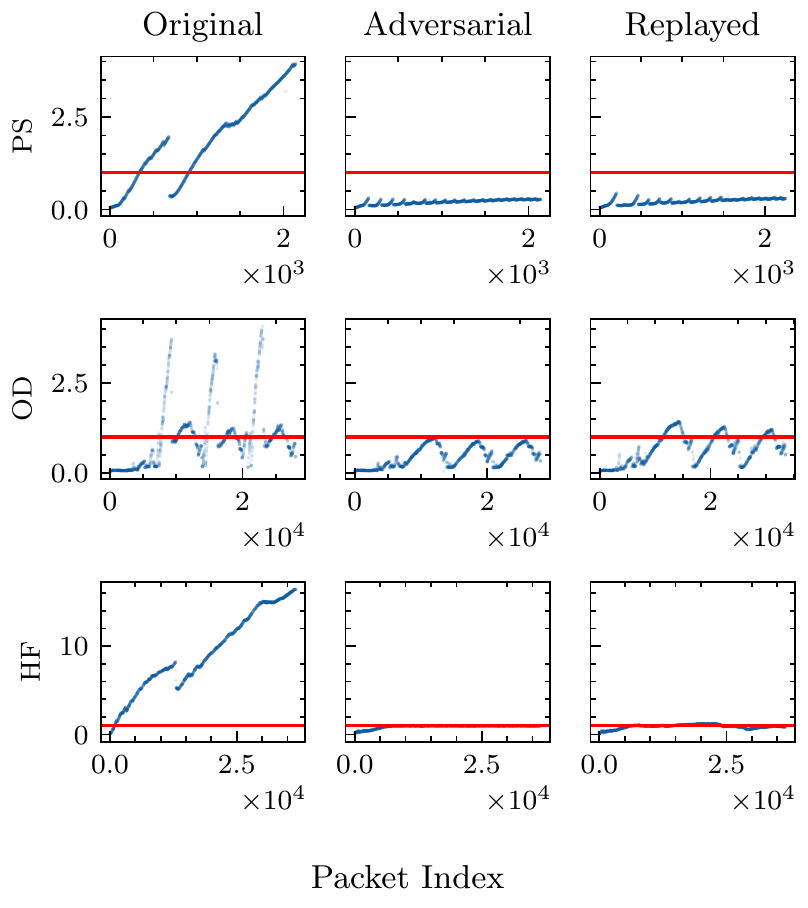}
    \caption{Comparison of Kitsune's anomaly scores between Original, Adversarial and Replayed traffic for Port Scan (PS), OS Detection (OD) and HTTP Flooding (HF). The anomaly scores are normalised with respect to the threshold (red line).}
    \label{fig:kitsune_as}
\end{figure}

\subsection{Evasiveness of Adversarial Traffic}
\label{sec:performance_analysis}
\subsubsection{Our Dataset}
We first measure the ability of adversarial traffic to bypass NIDS detection on our dataset by comparing the performance of various ML/DL based NIDS under the adversarial traffic generated by \textit{Liuer Mihou} compared to the original attack\footnote{We have tried to compare our work with that of Han \textit{et al.} \cite{han2020practical}. 
However, their open-source implementation did not show good results on our dataset (see Appendix \ref{sec:exp_han} for more details).}. Table \ref{tab:nids_aedr} shows the performance and evasion metrics of adversarial and replayed traffic on various NIDS. 

We have found several interesting aspects of the result from our evasiveness analysis. 
Firstly, \textbf{replayed traffic is less evasive compared to adversarial traffic}, indicated by RER being higher than AER for all attacks. 
By inspecting the adversarial and replay traffic, we have found that the inherent processing delays in the attacker's machine and propagation delays in the transmission medium cause the packets' arrival time at the victim to be slightly different from the arrival time specified in the adversarial traffic. 
Such delays will cause a slight increase in the adversarial packet's anomaly scores, potentially making the attack detectable. 
Nevertheless, the magnitude of the anomaly scores in both adversarial and replayed attacks have been reduced significantly, illustrated in Figure \ref{fig:kitsune_as}.

Secondly, \textbf{Kitsune is highly vulnerable to Liuer Mihou, but other anomaly detectors are not.} 
The ADR of \textit{Liuer Mihou} with Kitsune is 0 for all three attacks, while other detection algorithms have ADR higher than 0.
We have two hypothesises to explain this phenomenon. 
First, Kitsune is designed to have a relatively high threshold value compared to other domains to reduce false alarms.  
Even a seemingly low false positive rate of 0.01 can result in thousands of false alarms when millions of packets are processed in the NIDS domain. Hence, the high threshold value causes Kitsune to be more vulnerable to adversarial attacks.
Second, the internal feature representation of the surrogate model is not similar to other anomaly detectors. 
Different anomaly detectors have different design structures that favour different internal representations learned from benign traffic. 
The surrogate model used in our experiments is a vanilla autoencoder which has the same underlying detection algorithm as Kitsune, and they will learn similar feature representations. 
Other ML algorithms are not based on autoencoders, which means they will learn different features than the surrogate model and reduce the transferability of adversarial examples.

In order to gain more insight into the transferability of adversarial traffic, we conduct further experiments to measure the similarity of anomaly scores between the surrogate and target NIDS and the relative threshold values.

\begin{figure*}[t]
    \centering
    \includegraphics[width=\linewidth]{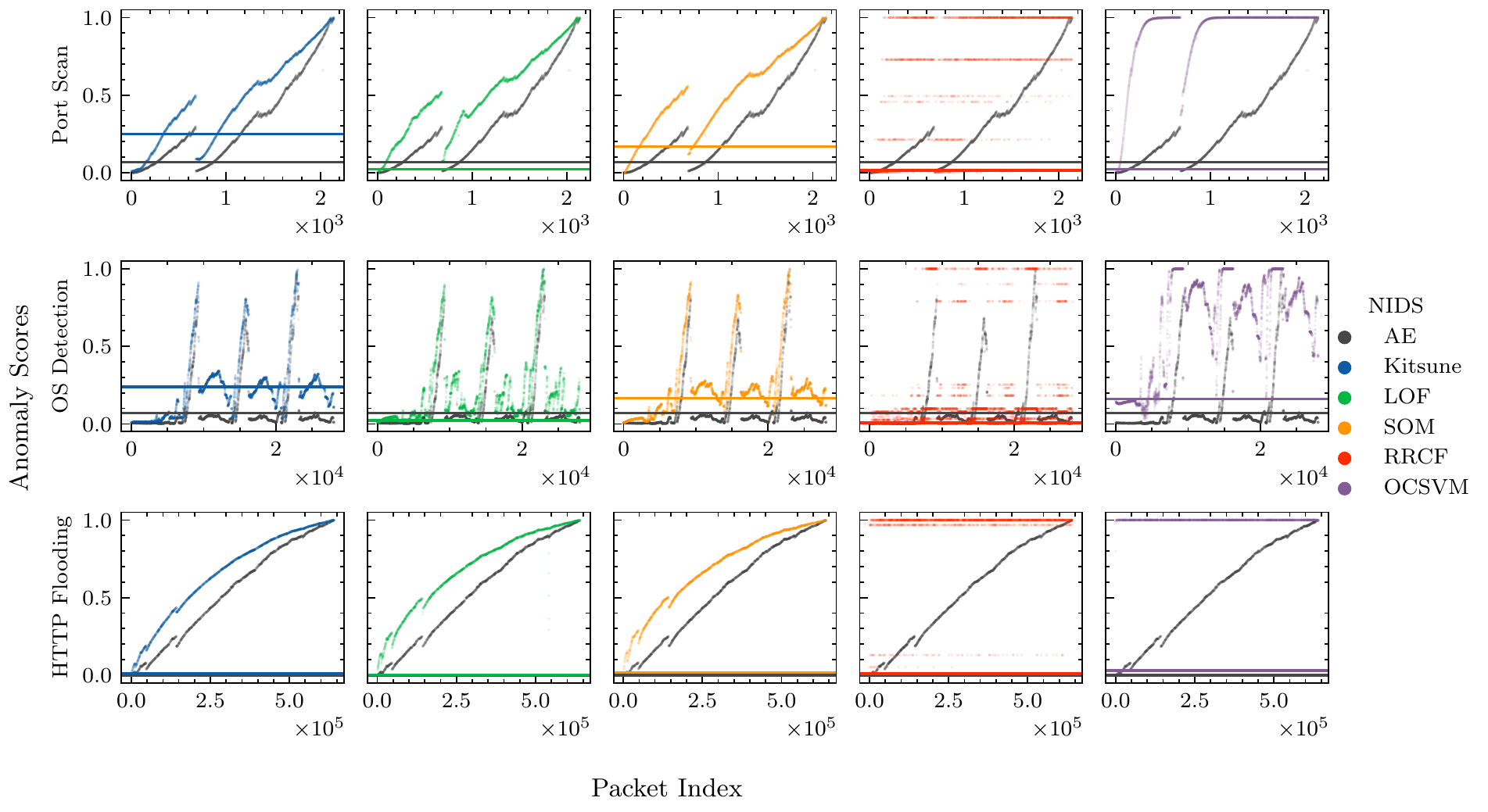}
    \caption{Anomaly scores of malicious traffic between the surrogate (orange) and various outlier detection algorithms (blue).}
    \label{fig:as_comp}
\end{figure*}

\begin{table}[t]
\caption{Adversarial detection and evasion rates of the adversarial traffic generated with Liuer Mihou against various NIDS.}
\label{tab:nids_aedr}
\begin{tabularx}{\linewidth}{@{}llYYYYYY@{}}
\toprule
\textbf{Attack} & \textbf{ML} & \textbf{TNR} & \textbf{MDR} & \textbf{ADR} & \textbf{RDR} & \textbf{AER} & \textbf{RER} \\ 
\midrule
PS & Kitsune & 1.000 & 0.74 & 0.00 & 0.00 & 1.00 & 1.00 \\
PS & SOM & 0.993 & 0.89 & 0.05 & 0.15 & 0.94 & 0.83 \\
PS & LOF & 0.999 & 0.98 & 0.98 & 0.98 & 0.00 & 0.00 \\
PS & RRCF & 0.992 & 0.93 & 0.71 & 0.71 & 0.24 & 0.23 \\
PS & OCSVM & 0.999 & 0.99 & 0.98 & 0.99 & 0.01 & 0.00 \\\rowcolor[HTML]{EFEFEF} 
OD & Kitsune & 1.000 & 0.41 & 0.00 & 0.25 & 1.00 & 0.39 \\\rowcolor[HTML]{EFEFEF} 
OD & SOM & 0.993 & 0.66 & 0.29 & 0.53 & 0.55 & 0.19 \\\rowcolor[HTML]{EFEFEF} 
OD & LOF & 0.999 & 0.95 & 0.95 & 0.97 & 0.00 & -0.01 \\\rowcolor[HTML]{EFEFEF} 
OD & RRCF & 0.995 & 0.64 & 0.46 & 0.55 & 0.28 & 0.15 \\\rowcolor[HTML]{EFEFEF} 
OD & OCSVM & 0.999 & 0.84 & 0.83 & 0.88 & 0.00 & -0.05 \\
HF & Kitsune & 1.000 & 1.00 & 0.00 & 0.33 & 1.00 & 0.67 \\
HF & SOM & 0.993 & 1.00 & 0.91 & 0.77 & 0.09 & 0.23 \\
HF & LOF & 0.999 & 1.00 & 1.00 & 1.00 & 0.00 & 0.00 \\
HF & RRCF & 0.978 & 1.00 & 1.00 & 1.00 & 0.00 & 0.00 \\
HF & OCSVM & 0.999 & 1.00 & 1.00 & 1.00 & 0.00 & 0.00 \\ \bottomrule
\end{tabularx}
\end{table}

\begin{table}[t]
\caption{Comparison of Euclidean Distance (ED) and threshold values ($\phi$) between the surrogate NIDS and different outlier detection algorithms.}
\label{tab:sim_res}
\begin{tabularx}{\linewidth}{@{}lXYYYY@{}}
\toprule
\textbf{Attack} & \textbf{Algorithm} & \textbf{ED} & $\boldsymbol{\phi_{alg}}$ & $\boldsymbol{\phi_{sur}}$ & $\boldsymbol{\Delta \phi}$ \\ \midrule

PS & Kitsune & 0.1416 & 0.2498 & 0.0684 & -0.1814 \\

PS & SOM & 0.1961 & 0.1696 & 0.0684 & -0.1011 \\

PS & LOF & 0.1780 & 0.0221 & 0.0684 & 0.0463 \\

PS & RRCF & 0.3990 & 0.0143 & 0.0684 & 0.0541 \\

PS & OCSVM & 0.4184 & 0.0214 & 0.0684 & 0.0470 \\
\rowcolor[HTML]{EFEFEF} 

OD & Kitsune & 0.1376 & 0.2398 & 0.0711 & -0.1687 \\
\rowcolor[HTML]{EFEFEF} 

OD & SOM & 0.1468 & 0.1663 & 0.0711 & -0.0952 \\
\rowcolor[HTML]{EFEFEF} 

OD & LOF & 0.1132 & 0.0217 & 0.0711 & 0.0494 \\
\rowcolor[HTML]{EFEFEF} 

OD & RRCF & 0.1598 & 0.0080 & 0.0711 & 0.0632 \\
\rowcolor[HTML]{EFEFEF} 

OD & OCSVM & 0.4417 & 0.1605 & 0.0711 & -0.0893 \\

HF & Kitsune & 0.1210 & 0.0096 & 0.0005 & -0.0091 \\

HF & SOM & 0.1527 & 0.0160 & 0.0005 & -0.0155 \\

HF & LOF & 0.1489 & 0.0018 & 0.0005 & -0.0013 \\

HF & OCSVM & 0.5405 & 0.0293 & 0.0005 & -0.0288 \\ 

HF & RRCF & 0.4329 & 0.0059 & 0.0005 & -0.0053 \\\bottomrule

\end{tabularx}
\end{table}

For a fair comparison between the models, we normalise the anomaly scores of the malicious attack and threshold values for each model to be between 0 and 1. 
The similarity between the target and surrogate NIDSes is measured by the average Euclidean Distance (ED) between each pair of anomaly scores. 
Table \ref{tab:sim_res} shows the ED and relative threshold difference between the autoencoder surrogate and various target NIDS, Figure \ref{fig:as_comp} illustrates the difference in anomaly scores.

Table \ref{tab:sim_res} reveals that Kitsune and LOF have low ED compared to the surrogate model, indicating they have learned similar decision functions. 
However, the relative threshold value of LOF is lower than the surrogate model, so that transferability is limited. 
On the other hand, Kitsune has a higher relative threshold value, and together with its high similarity, makes it highly vulnerable to autoencoder surrogates.

RRCF, SOM and OCSVM have a larger ED than the surrogate model mainly due to the design of the algorithm tends to learn different feature representations compared to autoencoders. 
For example, the tree structure of RRCF causes the output to be discrete, and SOM and OCSVM uses an RBF Kernel and Gaussian topology that makes the anomaly scores conform to Gaussian distribution. 
Therefore, the AER of adversarial traffic is low regardless of the difference threshold, and the adversarial traffic targeted at the surrogate model is not transferable to the target model.

\subsubsection{Kitsune Dataset}
In addition to our dataset, we measure the evasiveness of \textit{Liuer Mihou} on the Kitsune dataset \cite{mirsky2018kitsune}. 
We ran the attack with the same configuration described in Section \ref{sec:exp_setup}, and compare the reported AER of \textit{Liuer Mihou} to adversarial traffic generated with Traffic Manipulator with similar constraints \cite{han2020practical} (Traffic Manipulator have only reported AER for three attacks). 
Table \ref{tab:kitsune_exp} shows the ADR and AER of the adversarial traffic generated by \textit{Liuer Mihou} as well as the AER of Traffic Manipulator (TM).
The RDR and RER were not measured since we cannot replicate the same IoT testbed as Kitsune. 
Results show that most attacks have a high AER, indicating \textit{Liuer Mihou} can successfully evade Kitsune in general. 
In addition, our attack outperforms TM in Mirai and SSDP Flooding and is slightly less evasive for Fuzzing attacks.
However, none of the adversarial traffic can entirely bypass Kitsune, similar to what we have observed in our dataset.

\begin{table}[t]
\caption{The evasiveness of Liuer Mihou against Kitsune on the Kitsune dataset \cite{mirsky2018kitsune} compared to the AER of Traffic Manipulator (TM).}
\label{tab:kitsune_exp}
\begin{tabularx}{\linewidth}{@{}lYYY@{}}
\toprule
\textbf{Attacks} & \textbf{MDR}  & \textbf{AER-LM} & \textbf{AER-TM} \\ \midrule
Active Wiretap & 0.921 & 0.980 &  N/A\\
ARP MITM & 0.798  & 0.512 &  N/A\\
Fuzzing & 0.912 & 0.905 & \underline{0.955} \\
Mirai & 0.882 & \underline{0.823} & 0.721 \\
OS & 0.990 & 0.110 &  N/A\\
SSDP Flooding & 1.000  & \underline{0.791} & 0.532 \\
SSL Renegotiation & 0.987  & 0.450 &  N/A\\
SYN DoS & 0.379  & 0.974 &  N/A\\
Video Injection & 0.984  & 0.461 &  N/A \\ \bottomrule
\end{tabularx}
\end{table}

\begin{figure}[t]
    \centering
    \includegraphics[width=\linewidth]{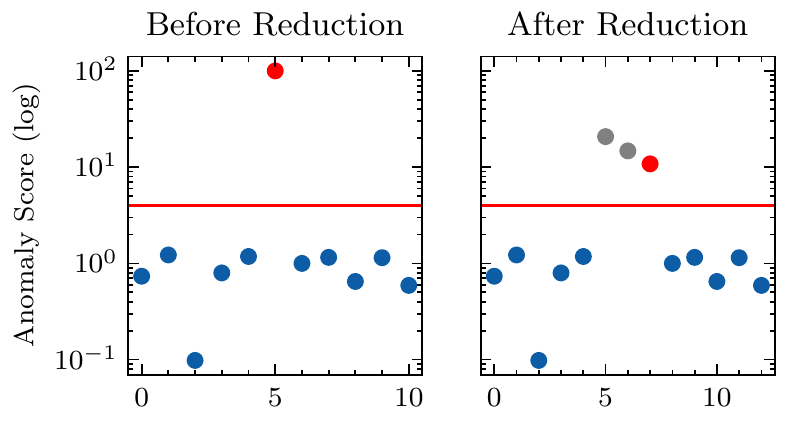}
    \caption{How \textit{Liuer Mihou} handles sudden spikes. Left: original traffic with sudden spike (red dot) above the threshold (red line). Right: the spike is reduced by injecting redundant packets (grey dots), but is still above the threshold.}
    \label{fig:aw_plot}
\end{figure}

Upon inspecting the Kitsune dataset in detail, we have found that the Kitsune dataset contains packets with abnormally high anomaly scores compared to the previous packet in all traffic files, creating sudden spikes in the traffic patterns. 
When there is a large spike in the anomaly score, \textit{Liuer Mihou} cannot reduce the anomaly score below the threshold with the predefined boundaries. However, it is still able to lower the anomaly scores of the malicious packet, illustrated in Figure \ref{fig:aw_plot}.
We suspect the large spikes in anomaly scores are caused by anonymisation. 
The authors of Kitsune \cite{mirsky2018kitsune} have truncated payload sizes to 200 bytes for privacy reasons, which results in a large number of malformed packets that are not representative of real IoT traffic.
Nevertheless, \textit{Liuer Mihou} can still increase the evasiveness of the attack, and the adversarial traffic can be made more evasive by recursively running \textit{Liuer Mihou} on adversarial traffic. \\[6pt]
\noindent \fbox{%
    \parbox{\linewidth - 2\fboxsep}{%
    \textbf{Summary.} The evasiveness of \textit{Liuer Mihou} largely depends on the similarity of learned representations and relative threshold values between the surrogate and target model.
    The evasiveness of the replayed attack will be lower due to inherent transmission and processing delays. 
    \textit{Liuer Mihou} is less effective when the attack contains large spikes in anomaly scores, but it can be mitigated by recursively running \textit{Liuer Mihou} on the adversarial attack.
    }%
}

\begin{table}[t]
\caption{Comparison between original and adversarial HTTP Flooding attacks.}
\label{tab:rep_flooding}
\begin{tabularx}{\linewidth}{@{}XYY@{}}
\toprule
\textbf{Traffic Type} & \textbf{RTT (ms)} & \textbf{RRD} \\ \midrule

\textit{Normal}    & 6.602    & 1.000          \\
\textit{Original} HF  & 173.393  & 26.264           \\
\textit{Liuer Mihou} HF  & 10.303   & 1.561          \\ \bottomrule
\end{tabularx}
\end{table}

\begin{table}[t]
\caption{Comparison between original and adversarial scanning attacks.}
\label{tab:rep_ps}
\newcolumntype{S}{>{\hsize=.5\hsize}X}
\newcolumntype{L}{>{\hsize=1.5\hsize}X}
\begin{tabularx}{\linewidth}{@{}lLSYY@{}}
\toprule
\textbf{Traffic} & \textbf{Ports Scanned} & \textbf{RPS} & \textbf{Time(s)} & \textbf{RTD} \\\midrule
\textit{Original} OD & 155 & 1 & 1,654.55 & 1.00 \\
\textit{Liuer Mihou} OD & 155 & 1 & 1,792.52 & 1.08 \\
\textit{Original} PS  & 155 & 1 & 1.99  & 1.00  \\
\textit{Liuer Mihou} PS & 150 & 0.968 & 19.99 & 10.04 \\ \bottomrule
\end{tabularx}
\end{table}

\subsection{Maliciousness of Adversary}
Under our threat model, the adversarial attack must also fully/partially maintain its original malicious functionality. 
We objectively compare the maliciousness of the replayed packets for each attack to the unmodified attack with malicious metrics with the metrics defined in Section \ref{sec:semantic_metrics}. 

The goal of HTTP Flooding is to make the Google Home device unresponsive, so we measure the average round trip time using ping, with results shown in Table \ref{tab:rep_flooding}. 
As expected, the delay in RTT caused by the replayed HTTP flooding attack is only a small proportion of the unmodified HTTP flooding, but it is still larger than normal. 
We have tried using Google Home during the adversarial flooding attacks, and from our experiences, the increase of RTT of the adversarial traffic generated with \textit{Liuer Mihou} was barely noticeable. 

Probing attacks, such as Port Scan and OS Detection attacks, aim to detect open ports of the device.
We compare the number of well-known ports (port number up to 1024) detected with the original attack to the adversarial attack and compare the relative time needed to get the results. 
Table \ref{tab:rep_ps} shows the attack metrics of OS Detection and Port Scan, respectively. 
Results show that adversarial OS Detection attacks can fully scan all the ports scanned by the unmodified attack with less than a ten percent increase in RTD for \textit{Liuer Mihou}. 
Port Scan attacks can scan over 90\% of the original ports but takes ten times more time than the original Port Scan attack. 

We hypothesise that the failure of adversarial HTTP flooding is mainly due to the maximum time delay being set very high, which causes the intensity to be overly reduced, and the adversarial attack is no longer malicious. Therefore, when generating adversarial traffic for DoS attacks, the maximum time delay of a packet should be carefully set to preserve its maliciousness.
On the other hand, probing attacks rely on the payload content to conduct malicious activities, and the maximum time delay of a packet will not significantly impact the malicious functionalities.
Another contributing factor to maliciousness is the surrogate threshold. Intuitively, the surrogate threshold measures how strict the traffic has to conform to the learned representation of normality. Since the surrogate model has a relatively low threshold, it forces the adversarial traffic to conform to a stricter notion of normality and potentially remove DoS attacks' maliciousness. Interestingly, probing attacks did not have much reduction in maliciousness with the same threshold. This is because the features extracted by Kitsune focus heavily on packet inter-arrival times and are most sensitive to inter-arrival times. Hence, probing attacks are detected not because of the payload contents but because the inter-arrival times are shorter than usual (this also explains why Kitsune cannot detect ARP spoofing and poisoning since the two attacks transmit packets at a low frequency). Since the probing attacks' inter-arrival time has no significant effect on the maliciousness, the adversarial probing attacks are still malicious.   
\\[6pt]

\noindent \fbox{%
    \parbox{\linewidth - 2\fboxsep}{%
    \textbf{Summary.} 
    The maximum time delay can significantly affect the maliciousness of DoS attacks that relies on packet arrival time. Therefore, the maximum time delay should be carefully set to retain the malicious functionality. 
    The maliciousness of probing attacks that mainly relies on packet payload is less affected by the mutation operations. 
    }%
}

\subsection{Evasiveness against Defences}
\label{sec:defence_analysis}
We evaluate the strength of \textit{Liuer Mihou} against Kitsune with adversarial defence mechanisms deployed because it is most vulnerable. 
Common adversarial defences include adversarial training \cite{madry2017towards}, where the model is trained with adversarial examples with correct labels to increase its robustness, and adversarial detection \cite{xu2017feature, meng2017magnet}, where a secondary classifier is trained to detect adversarial examples. Although adversarial training have shown great potential in defending against adversarial attacks, it is not applicable under our threat model, because our NIDS is trained in an unsupervised manner without any label information. 
Thus, we choose two adversarial detection defences that are model and domain agnostic: Feature Squeezing \cite{xu2017feature}, and Mag-Net \cite{meng2017magnet}. 
With the addition of adversarial defences, the adversarial defences will first determine whether the input feature is adversarial or clean. 
Next, Kitsune will detect whether the input example is benign or malicious.

\begin{table}[t]
\caption{Detection results of Kitsune with Feature Squeezing.}
\label{tab:fs_conf_mat}
\begin{tabularx}{\linewidth}{@{}lYY YYY@{}}
\toprule
 & \multicolumn{2}{c}{\textbf{FS Output}} & \multicolumn{2}{c}{\textbf{Kitsune Output}} & \\ \cmidrule(lr){2-3} \cmidrule(lr){4-5}
\textbf{Traffic} & \textbf{Clean} & \textbf{Adv.} & \textbf{Benign} & \textbf{Malicious} & \textbf{Total} \\ \midrule

Benign & 14400 & 0 & 14400 & 0 & 14400 \\

$\textrm{PS}_{\textrm{mal}}$ & 1975 & 167 & 565 & 1577 & 2142 \\

$\textrm{PS}_{\textrm{adv}}$ & 2142 & 0 & 2142 & 0 & 2142 \\

$\textrm{PS}_{\textrm{rep}}$ & 2252 & 0 & 2252 & 0 & 2252 \\

$\textrm{OD}_{\textrm{mal}}$ & 27075 & 949 & 16651 & 11373 & 28024 \\

$\textrm{OD}_{\textrm{adv}}$ & 28074 & 0 & 28074 & 0 & 28074 \\

$\textrm{OD}_{\textrm{rep}}$ & 33571 & 0 & 33571 & 8313 & 33571 \\

$\textrm{HF}_{\textrm{mal}}$ & 33462 & 607305 & 935 & 639832 & 640767 \\

$\textrm{HF}_{\textrm{adv}}$ & 37440 & 0 & 37440 & 0 & 37440 \\

$\textrm{HF}_{\textrm{rep}}$ & 36653 & 0 & 36653 & 12089 & 36653 \\ \bottomrule
\end{tabularx}
\end{table}
\begin{table}[t]
\caption{Detection results of Kitsune with Mag-Net.}
\label{tab:mn_conf_mat}
\begin{tabularx}{\linewidth}{@{}lYY YYY@{}}
\toprule
 & \multicolumn{2}{c}{\textbf{Mag-Net Output}} & \multicolumn{2}{c}{\textbf{Kitsune Output}} &  \\ \cmidrule(lr){2-3} \cmidrule(lr){4-5}
\textbf{Traffic} & \textbf{Clean} & \textbf{Adv.} & \textbf{Benign} & \textbf{Malicious} & \textbf{Total} \\ \midrule

Benign & 14400 & 0 & 14400 & 0 & 14400 \\

$\textrm{PS}_{\textrm{mal}}$ & 126 & 2016 & 2142 & 0 & 2142 \\

$\textrm{PS}_{\textrm{adv}}$ & 750 & 1392 & 2142 & 0 & 2142 \\

$\textrm{PS}_{\textrm{rep}}$ & 702 & 1550 & 2252 & 0 & 2252 \\

$\textrm{OD}_{\textrm{mal}}$ & 7400 & 20624 & 27789 & 235 & 28024 \\

$\textrm{OD}_{\textrm{adv}}$ & 9662 & 18412 & 28074 & 0 & 28074 \\

$\textrm{OD}_{\textrm{rep}}$ & 7674 & 25897 & 33571 & 0 & 33571 \\

$\textrm{HF}_{\textrm{mal}}$ & 322 & 640445 & 640118 & 649 & 640767 \\

$\textrm{HF}_{\textrm{adv}}$ & 690 & 36750 & 37440 & 0 & 37440 \\

$\textrm{HF}_{\textrm{rep}}$ & 728 & 35925 & 36653 & 0 & 36653 \\ \bottomrule
\end{tabularx}
\end{table}

\subsubsection{Feature Squeezing} 
Feature Squeezing (FS) \cite{xu2017feature} aims to reduce the precision of the input feature space and limit the degree of perturbations available to the attacker. 
The general strategy is to compare the classifier's prediction between original and squeezed inputs, and if there is a significant difference in prediction, the sample is alerted as adversarial. 

We modify Kitsune to squeeze the features with four precision levels, $1,2,3,4$ decimal places, and a threshold value is based on the absolute difference between the unsqueezed and squeezed data on benign traffic for each precision level. 
Table \ref{tab:fs_conf_mat} shows the detection result of FS and Kitsune on benign, malicious, adversarial and replayed traffic. 
Results show FS fails to classify any adversarial and replay traffic as adversarial. 
Instead, it classifies malicious packets as adversarial, which suggest FS is not a valid defence against \textit{Liuer Mihou}.

\subsubsection{Mag-Net} 
Mag-Net \cite{meng2017magnet} utilises a group of detectors and a reformer to remove any adversarial perturbation. 
The detectors perform a preliminary check that determines whether the input sample looks clean and raise an alert if it does not. 
Next, all inputs, regardless of the detector's classification, are passed to the reformer that reconstructs the input to remove any adversarial perturbation.
The final detection is performed on the reconstructed input. 

We have used the two default detectors provided by the GitHub implementation of Mag-Net \cite{magnet_github}. 
We first train the two detectors and the reformer on benign traffic and set the detector's threshold as the maximum RMSE value on benign traffic. 
Next, we record the output by Mag-Net's detector and the output of Kitsune on reformed input for benign, malicious, adversarial and replayed traffic. 
Table \ref{tab:mn_conf_mat} presents the experiment result. 
From the table, we notice that detectors of Mag-Net have detected both malicious and adversarial traffic as adversarial. 
However, the reconstruction by the reformer removes the malicious and adversarial characteristics of the feature, which causes Kitsune to classify all traffic as benign. 
Therefore, Mag-Net is also not a suitable adversarial defence for NIDS. \\[6pt]

\noindent \fbox{%
    \parbox{\linewidth - 2\fboxsep}{%
    \textbf{Summary.} 
    Feature-level adversarial detectors are not applicable for NIDS because adversarial attack in NIDS makes packet-level modifications that can potentially have a large difference compared to the original attack. Moreover, the detectors are trained on purely benign data, which does not allow the detectors to distinguish between adversarial and malicious traffic. 
    }%
}
\section{Related Work}
\label{sec:background}

\subsection{Anomaly-based NIDS}
\label{sec:kitsune}
The target NIDS of \textit{Liuer Mihou} is anomaly-based NIDS, where an anomaly detection algorithm is deployed as its detection engine as opposed to classification. 
The advantage of using anomaly detection instead of classification is that they are unsupervised and do not require any labelled data, which is more suitable for NIDS since labelling millions of packets each day is not feasible.

The general process of anomaly-based NIDS comprises three main stages: packet capturing, feature extraction and anomaly detection.
When a packet arrives, the NIDS uses packet capturing libraries (\textit{e.g.,} Scapy and tshark) to capture the packets. 
The packets are passed on to the feature extractor, which extracts a wide range of aggregate features about the packets. 
To allow fast extraction of features, common feature extractors \cite{sharafaldin2018toward, mirsky2018kitsune} extract statistics provided in the header such as payload and arrival time without inspecting the payload.
    
    
    

The extracted features form the input to the anomaly detector.
The output of the anomaly detector is often a number indicating its normality, which we will refer to as the anomaly score of the packet. 
During training, the NIDS calculates a suitable threshold based on anomaly scores of benign data with a predefined heuristic. For example, using the maximum anomaly score or three standard deviations away from the mean. 
During execution, the NIDS classifies any packets with an anomaly score above the threshold as malicious.

\subsection{Adversarial Attacks bypassing NIDS}
\label{sec:taxonomy}
Existing adversarial attacks in the NIDS domain typically alter the input in the feature space, \textit{i.e.,} assume a white-box scenario and apply adversarial attacks designed for images (\textit{e.g.,} FGSM \cite{goodfellow2014explaining}, JSMA \cite{papernot2016limitations}, C \& W attack \cite{carlini2017towards}, and DeepFool \cite{moosavi2016deepfool}) to generated adversarial network features. 
However, having just the adversarial network features have no practical significance as the features alone cannot be used to conduct any malicious network attacks. 
Hence, problem space attacks \cite{pierazzi2020intriguing} are needed to enhance the practicality of adversarial attacks.

A handful of studies have considered adversarial attacks that modify the raw packets instead of network features in the NIDS domain. 
Overall, there are two main approaches to packet-level modification. 
One method is to predefine a set of packet-level mutations at various levels, for example, fragmenting the packet or delaying the packet. 
Then randomly apply the mutation and test if it can bypass detection \cite{homoliak2018improving, hashemi2019towards}. 
These methods require vigorous trial-and-error and provide little to no theoretical guidance and insight.
A more formal approach formulates the adversarial attack as a bi-level optimisation problem. These attacks first generate realistic adversarial features with Generative Adversarial Networks \cite{han2020practical} or Manifold Approximation \cite{kuppa2019black}. 
Next, the packets are modified to minimise the distance between the adversarial and malicious features.
However, the bi-level optimisation formulation complicates the search space and is difficult for the search algorithm to find a solution.

\subsection{Adversarial Defences}
\label{defenses}
Adversarial attacks have urged researchers to develop countermeasures to detect or mitigate the effect of adversarial examples. 
Multiple defence mechanisms and paradigms have been proposed targeting classifiers in CV. 
However, most of them can be broken by adaptive adversaries that have knowledge of the defence mechanisms in place \cite{carlini2017adversarial, athalye2018obfuscated}. 
The most effective and promising defence mechanisms are adversarial training \cite{madry2017towards} and adversarial detection  \cite{meng2017magnet, xu2017feature}. 
Adversarial training aims to smooth out discontinuities in the feature space to correctly classify the adversarial examples, which requires retraining the model. 
In contrast, adversarial detection recognises that adversarial examples are synthetically created and contain different characteristics than clean examples. Hence, adversarially perturbed examples can be effectively detected.

Most adversarial defences are evaluated in the CV domain and on classification algorithms.
To the best of our knowledge, we could not find any adversarial defence explicitly designed for NIDS, possibly due to the lack of practical adversarial attacks that are available in the first place. 
Furthermore, CV and NIDS have drastically different feature space structures, and CV often uses supervised learning algorithms, whereas NIDS uses unsupervised algorithms. Hence, it is an open question whether existing adversarial defences designed for CV can be directly deployed in the NIDS domain.

\section{Discussion}
\label{sec:discussion}

\subsection{Evasiveness and Maliciousness Tradeoff}
\textit{Liuer Mihou} is a transfer-based attack that leverages adversarial transferability across ML/DL algorithms. 
Adversarial transferability has been studied mainly with regards to supervised, multi-class classification \cite{papernot2016transferability, adam2018stochastic, pang2019improving}, but for unsupervised binary classification problems such as NIDS, it is a mostly untouched topic. 

From our experiment results in Section \ref{sec:performance_analysis}, we have empirically shown that adversarial transferability across anomaly detection based NIDS largely depends on the similarity of the decision boundary and relative threshold value between the target and surrogate NIDSes. 
The similarity of the decision boundary ensures the packet-level modifications on the surrogate have the same effect on the target and a lower threshold value ensures the surrogate forms a stricter benign profile. Since knowing the actual detection model is infeasible in practice, the similarity of the decision boundary cannot be increased. Therefore, to increase evasiveness, the attacker have to rely on lowering the threshold value of the surrogate model.

Analysis of the replayed adversarial traffic shows that the replayed packets’ anomaly scores will be slightly higher than the theoretically generated adversarial packets due to inherent propagation delays and process delays that cause the packets to arrive at slightly different times. Therefore, surrogate NIDS’s threshold is encouraged to be low and create a buffer zone for inherent delays. 

For the adversarial and replayed attack to be evasive, the surrogate threshold is encouraged to be low. However, the surrogate threshold represents how similar the packet is to the normal traffic, and lowering the surrogate threshold will inevitably lower the maliciousness of the adversarial traffic. For example, our adversarial HTTP Flooding did not significantly increase the target device’s response time, indicating that the threshold value of the surrogate model is too low to preserve the maliciousness of HTTP Flooding attacks. Hence, adversarial attacks on NIDS face a fundamental trade-off between evasiveness and maliciousness that can be adjusted via the surrogate threshold. It is crucial to find a balance between them to conduct successful adversarial attacks.

\subsection{Weakness of Adversarial Defence}
We have evaluated two plug-and-play adversarial defence methods: Feature Squeezing \cite{xu2017feature} and Mag-Net \cite{meng2017magnet}. 
These defences were designed initially for classification algorithms, and the results from our experiment in Section \ref{sec:defence_analysis} have shown that both methods are unsuitable in the NIDS domain. 

There are two main research challenges for applying adversarial detectors in CV to NIDS.
First, the design principle of adversarial detectors relies on the fact that adversarial features are synthetically created and have small distributional differences from clean, natural input. 
However, packet-level attacks such as \textit{Liuer Mihou} modify the packets directly rather than the features, and feature-level changes do not have to be minimal as long as the maliciousness of the original attack is preserved. 
Hence, packet-level attacks make large changes in the input feature and have realistic distribution as benign traffic, and adversarial detectors that intentionally ignore small perturbations, such as Feature Squeezing, fail to detect the adversarial examples.
Second, only benign traffic is available to train the adversarial detectors under a realistic NIDS threat model. 
Therefore, the adversarial detector cannot distinguish between adversarial and malicious traffic, classifying all attacks as adversarial. 
In essence, the adversarial detector becomes another NIDS, which is redundant.
Some adversarial detectors, such as Mag-Net, attempts to remove adversarial perturbation. 
Since the detector cannot distinguish between whether the perturbation comes from malicious or adversarial traffic, it will remove both types of perturbations and make all inputs benign, which helps the attack evade detection.

Due to the limitation of adversarial detectors in the NIDS setting, new forms of adversarial defence have to be developed.
Our experiment results in Section \ref{sec:performance_analysis} shown \textit{Liuer Mihou} relies on adversarial transferability to be successful, which requires training a suitable surrogate NIDS to mimic the target NIDS. 
Therefore, to block adversarial transferability the defender can introduce randomness in the network by stochastically changing the NIDS structure \cite{sengupta2019mtdeep} to force the attacker to generate universal adversarial examples against a wide range of detection algorithms.

\subsection{Limitations and Future Work}
\label{limitations}
We evaluate \textit{Liuer Mihou} under a grey-box scenario where the attacker is assumed to have complete knowledge of the feature extractor. 
Although these assumptions are justified and aligned with previous works, future work could investigate the performance of \textit{Liuer Mihou} under a black-box threat model with a surrogate feature extractor.

The number of mutation operations we have defined is limited and rather simple. 
Future studies could expand the mutation operations, such as fragmentation and reordering, to generate more evasive adversarial examples. 
Moreover, the redundant packets were set to have the same type as malicious packets with randomly generated payloads. 
For attacks that depend heavily on the payload content, poorly crafted packets may result in unexpected behaviour of the adversarial traffic. 
Future studies can investigate ways to design redundant packets that are more evasive and investigate the effect of mutation operations on the extracted features. 

We have only evaluated the maliciousness of \textit{Liuer Mihou} on a limited number of network attacks and IoT devices. 
Future work can extend the number of network attacks to include Man-in-the-Middle, brute force, and fuzzing. The target device can also include broader range of IoT devices, such as security cameras, smart lightbulbs, and smart TVs.

Another limitation of our work is that we have evaluated \textit{Liuer Mihou} in a rather simple IoT testbed that assumes no packets are dropped and devices are not congested. 
Future work could consider evaluating our attack under complicated network environments where high packet loss rate and processing delay are observed. 

\section{Conclusion}
\label{sec:conclusion}
We have proposed a novel, practical adversarial attack targeted at NIDS called \textit{Liuer Mihou}. 
The attack leverages adversarial transferability to train a surrogate NIDS to mimic the decision boundary of the target NIDS. 
The adversarial packets are generated by finding a set of mutation operations that minimises the anomaly score produced by the surrogate NIDS, which is transferable to the target NIDS. 

We have evaluated the evasiveness of the adversarial traffic against four ML-based algorithms (SOM, RRCF, LOF, and OCSVM) and the state-of-the-art IoT NIDS, Kitsune. The results of our experiments show \textit{Liuer Mihou} will be highly evasive if the surrogate NIDS has similar decision boundaries and a relatively low threshold compared to the target NIDS. 
The maliciousness of our attack is largely affected by the boundaries of mutation operation and the surrogate threshold, which leads to the trade-off between evasiveness and maliciousness. Hence the boundaries of mutation operations and surrogate threshold should be set carefully to ensure the original maliciousness is retained.
Finally, we demonstrated that adversarial detection defences such as Feature Squeezing and Mag-Net could not defend against \textit{Liuer Mihou} and were unsuitable for NIDS due to the lack of malicious and adversarial data available during training and that \textit{Liuer Mihou} can create a significant difference in feature space compared to the original feature. 

Our work provides a solid theoretical foundation for generating transfer-based practical adversarial traffic against NIDS and provides insightful discussion on adversarial transferability and defences in the NIDS domain.

\bibliographystyle{ACM-Reference-Format}
\bibliography{bib}


\begin{thebibliography}{00}


\ifx \showCODEN    \undefined \def \showCODEN     #1{\unskip}     \fi
\ifx \showDOI      \undefined \def \showDOI       #1{#1}\fi
\ifx \showISBNx    \undefined \def \showISBNx     #1{\unskip}     \fi
\ifx \showISBNxiii \undefined \def \showISBNxiii  #1{\unskip}     \fi
\ifx \showISSN     \undefined \def \showISSN      #1{\unskip}     \fi
\ifx \showLCCN     \undefined \def \showLCCN      #1{\unskip}     \fi
\ifx \shownote     \undefined \def \shownote      #1{#1}          \fi
\ifx \showarticletitle \undefined \def \showarticletitle #1{#1}   \fi
\ifx \showURL      \undefined \def \showURL       {\relax}        \fi
\providecommand\bibfield[2]{#2}
\providecommand\bibinfo[2]{#2}
\providecommand\natexlab[1]{#1}
\providecommand\showeprint[2][]{arXiv:#2}

\bibitem[\protect\citeauthoryear{Adam, Smirnov, Duvenaud, Haibe-Kains, and
  Goldenberg}{Adam et~al\mbox{.}}{2018}]%
        {adam2018stochastic}
\bibfield{author}{\bibinfo{person}{George~A Adam}, \bibinfo{person}{Petr
  Smirnov}, \bibinfo{person}{David Duvenaud}, \bibinfo{person}{Benjamin
  Haibe-Kains}, {and} \bibinfo{person}{Anna Goldenberg}.}
  \bibinfo{year}{2018}\natexlab{}.
\newblock \showarticletitle{Stochastic combinatorial ensembles for defending
  against adversarial examples}.
\newblock \bibinfo{journal}{{\em arXiv preprint arXiv:1808.06645\/}}
  (\bibinfo{year}{2018}).
\newblock


\bibitem[\protect\citeauthoryear{Al-Garadi, Mohamed, Al-Ali, Du, Ali, and
  Guizani}{Al-Garadi et~al\mbox{.}}{2020}]%
        {al2020survey}
\bibfield{author}{\bibinfo{person}{Mohammed~Ali Al-Garadi},
  \bibinfo{person}{Amr Mohamed}, \bibinfo{person}{Abdulla~Khalid Al-Ali},
  \bibinfo{person}{Xiaojiang Du}, \bibinfo{person}{Ihsan Ali}, {and}
  \bibinfo{person}{Mohsen Guizani}.} \bibinfo{year}{2020}\natexlab{}.
\newblock \showarticletitle{A survey of machine and deep learning methods for
  internet of things (IoT) security}.
\newblock \bibinfo{journal}{{\em IEEE Communications Surveys \& Tutorials\/}}
  \bibinfo{volume}{22}, \bibinfo{number}{3} (\bibinfo{year}{2020}),
  \bibinfo{pages}{1646--1685}.
\newblock


\bibitem[\protect\citeauthoryear{Anonymous}{Anonymous}{2020}]%
        {kudataset}
\bibfield{author}{\bibinfo{person}{Anonymous}.}
  \bibinfo{year}{2020}\natexlab{}.
\newblock \bibinfo{title}{IoT intrusion detection dataset}.
\newblock \bibinfo{howpublished}{{to be disclosed later}}.
  (\bibinfo{year}{2020}).
\newblock


\bibitem[\protect\citeauthoryear{Anonymous}{Anonymous}{2021}]%
        {lm_github}
\bibfield{author}{\bibinfo{person}{Anonymous}.}
  \bibinfo{year}{2021}\natexlab{}.
\newblock \bibinfo{title}{Anonymous}.
\newblock
  \bibinfo{howpublished}{\url{https://github.com/XXXX-5/automatic-waddle}}.
  (\bibinfo{year}{2021}).
\newblock
\newblock
\shownote{Intentionally anonymous.}


\bibitem[\protect\citeauthoryear{Appneta}{Appneta}{2020}]%
        {tcpreplay}
\bibfield{author}{\bibinfo{person}{Appneta}.} \bibinfo{year}{2020}\natexlab{}.
\newblock \bibinfo{title}{Tcpreplay - Pcap editing and replaying utilities}.
\newblock \bibinfo{howpublished}{\url{https://tcpreplay.appneta.com/}}.
  (\bibinfo{year}{2020}).
\newblock
\newblock
\shownote{Accessed: 2020-9-15.}


\bibitem[\protect\citeauthoryear{Arp, Quiring, Pendlebury, Warnecke, Pierazzi,
  Wressnegger, Cavallaro, and Rieck}{Arp et~al\mbox{.}}{2020}]%
        {arp2020and}
\bibfield{author}{\bibinfo{person}{Daniel Arp}, \bibinfo{person}{Erwin
  Quiring}, \bibinfo{person}{Feargus Pendlebury}, \bibinfo{person}{Alexander
  Warnecke}, \bibinfo{person}{Fabio Pierazzi}, \bibinfo{person}{Christian
  Wressnegger}, \bibinfo{person}{Lorenzo Cavallaro}, {and}
  \bibinfo{person}{Konrad Rieck}.} \bibinfo{year}{2020}\natexlab{}.
\newblock \showarticletitle{Dos and Don'ts of Machine Learning in Computer
  Security}.
\newblock \bibinfo{journal}{{\em arXiv preprint arXiv:2010.09470\/}}
  (\bibinfo{year}{2020}).
\newblock


\bibitem[\protect\citeauthoryear{Athalye, Carlini, and Wagner}{Athalye
  et~al\mbox{.}}{2018}]%
        {athalye2018obfuscated}
\bibfield{author}{\bibinfo{person}{Anish Athalye}, \bibinfo{person}{Nicholas
  Carlini}, {and} \bibinfo{person}{David Wagner}.}
  \bibinfo{year}{2018}\natexlab{}.
\newblock \showarticletitle{Obfuscated gradients give a false sense of
  security: Circumventing defenses to adversarial examples}.
\newblock \bibinfo{journal}{{\em arXiv preprint arXiv:1802.00420\/}}
  (\bibinfo{year}{2018}).
\newblock


\bibitem[\protect\citeauthoryear{Bartos, Mullapudi, and Troutman}{Bartos
  et~al\mbox{.}}{2019}]%
        {bartos_2019_rrcf}
\bibfield{author}{\bibinfo{person}{Matthew Bartos}, \bibinfo{person}{Abhiram
  Mullapudi}, {and} \bibinfo{person}{Sara Troutman}.}
  \bibinfo{year}{2019}\natexlab{}.
\newblock \showarticletitle{Implementation of the Robust Random Cut Forest
  algorithm for anomaly detection on streams}.
\newblock \bibinfo{journal}{{\em Journal of Open Source Software\/}}
  \bibinfo{volume}{4}, \bibinfo{number}{35} (\bibinfo{year}{2019}),
  \bibinfo{pages}{1336}.
\newblock


\bibitem[\protect\citeauthoryear{Blum and Roli}{Blum and Roli}{2003}]%
        {blum2003metaheuristics}
\bibfield{author}{\bibinfo{person}{Christian Blum} {and}
  \bibinfo{person}{Andrea Roli}.} \bibinfo{year}{2003}\natexlab{}.
\newblock \showarticletitle{Metaheuristics in combinatorial optimization:
  Overview and conceptual comparison}.
\newblock \bibinfo{journal}{{\em ACM computing surveys (CSUR)\/}}
  \bibinfo{volume}{35}, \bibinfo{number}{3} (\bibinfo{year}{2003}),
  \bibinfo{pages}{268--308}.
\newblock


\bibitem[\protect\citeauthoryear{Bochkovskiy, Wang, and Liao}{Bochkovskiy
  et~al\mbox{.}}{2020}]%
        {bochkovskiy2020yolov4}
\bibfield{author}{\bibinfo{person}{Alexey Bochkovskiy},
  \bibinfo{person}{Chien-Yao Wang}, {and} \bibinfo{person}{Hong-Yuan~Mark
  Liao}.} \bibinfo{year}{2020}\natexlab{}.
\newblock \showarticletitle{YOLOv4: Optimal Speed and Accuracy of Object
  Detection}.
\newblock \bibinfo{journal}{{\em arXiv preprint arXiv:2004.10934\/}}
  (\bibinfo{year}{2020}).
\newblock


\bibitem[\protect\citeauthoryear{Carlini and Wagner}{Carlini and
  Wagner}{2017a}]%
        {carlini2017adversarial}
\bibfield{author}{\bibinfo{person}{Nicholas Carlini} {and}
  \bibinfo{person}{David Wagner}.} \bibinfo{year}{2017}\natexlab{a}.
\newblock \showarticletitle{Adversarial examples are not easily detected:
  Bypassing ten detection methods}. In \bibinfo{booktitle}{{\em Proceedings of
  the 10th ACM Workshop on Artificial Intelligence and Security}}.
  \bibinfo{pages}{3--14}.
\newblock


\bibitem[\protect\citeauthoryear{Carlini and Wagner}{Carlini and
  Wagner}{2017b}]%
        {carlini2017towards}
\bibfield{author}{\bibinfo{person}{Nicholas Carlini} {and}
  \bibinfo{person}{David Wagner}.} \bibinfo{year}{2017}\natexlab{b}.
\newblock \showarticletitle{Towards evaluating the robustness of neural
  networks}. In \bibinfo{booktitle}{{\em 2017 ieee symposium on security and
  privacy (sp)}}. IEEE, \bibinfo{pages}{39--57}.
\newblock


\bibitem[\protect\citeauthoryear{Clements, Yang, Sharma, Hu, and Lao}{Clements
  et~al\mbox{.}}{2019}]%
        {clements2019rallying}
\bibfield{author}{\bibinfo{person}{Joseph Clements}, \bibinfo{person}{Yuzhe
  Yang}, \bibinfo{person}{Ankur Sharma}, \bibinfo{person}{Hongxin Hu}, {and}
  \bibinfo{person}{Yingjie Lao}.} \bibinfo{year}{2019}\natexlab{}.
\newblock \showarticletitle{Rallying adversarial techniques against deep
  learning for network security}.
\newblock \bibinfo{journal}{{\em arXiv preprint arXiv:1903.11688\/}}
  (\bibinfo{year}{2019}).
\newblock


\bibitem[\protect\citeauthoryear{Dorigo, Birattari, and Stutzle}{Dorigo
  et~al\mbox{.}}{2006}]%
        {dorigo2006ant}
\bibfield{author}{\bibinfo{person}{Marco Dorigo}, \bibinfo{person}{Mauro
  Birattari}, {and} \bibinfo{person}{Thomas Stutzle}.}
  \bibinfo{year}{2006}\natexlab{}.
\newblock \showarticletitle{Ant colony optimization}.
\newblock \bibinfo{journal}{{\em IEEE computational intelligence magazine\/}}
  \bibinfo{volume}{1}, \bibinfo{number}{4} (\bibinfo{year}{2006}),
  \bibinfo{pages}{28--39}.
\newblock


\bibitem[\protect\citeauthoryear{Engelbrecht}{Engelbrecht}{2013}]%
        {engelbrecht2013particle}
\bibfield{author}{\bibinfo{person}{Andries~Petrus Engelbrecht}.}
  \bibinfo{year}{2013}\natexlab{}.
\newblock \showarticletitle{Particle swarm optimization: Global best or local
  best?}. In \bibinfo{booktitle}{{\em 2013 BRICS congress on computational
  intelligence and 11th Brazilian congress on computational intelligence}}.
  IEEE, \bibinfo{pages}{124--135}.
\newblock


\bibitem[\protect\citeauthoryear{Goodfellow, Shlens, and Szegedy}{Goodfellow
  et~al\mbox{.}}{2014}]%
        {goodfellow2014explaining}
\bibfield{author}{\bibinfo{person}{Ian~J Goodfellow}, \bibinfo{person}{Jonathon
  Shlens}, {and} \bibinfo{person}{Christian Szegedy}.}
  \bibinfo{year}{2014}\natexlab{}.
\newblock \showarticletitle{Explaining and harnessing adversarial examples}.
\newblock \bibinfo{journal}{{\em arXiv preprint arXiv:1412.6572\/}}
  (\bibinfo{year}{2014}).
\newblock


\bibitem[\protect\citeauthoryear{Google}{Google}{2020}]%
        {tf2anomaly}
\bibfield{author}{\bibinfo{person}{Google}.} \bibinfo{year}{2020}\natexlab{}.
\newblock \bibinfo{title}{Intro to Autoencoders}.
\newblock
  \bibinfo{howpublished}{\url{https://www.tensorflow.org/tutorials/generative/autoencoder\#third_example_anomaly_detection}}.
    (\bibinfo{year}{2020}).
\newblock
\newblock
\shownote{Accessed: 2020-9-15.}


\bibitem[\protect\citeauthoryear{Guha, Mishra, Roy, and Schrijvers}{Guha
  et~al\mbox{.}}{2016}]%
        {guha2016robust}
\bibfield{author}{\bibinfo{person}{Sudipto Guha}, \bibinfo{person}{Nina
  Mishra}, \bibinfo{person}{Gourav Roy}, {and} \bibinfo{person}{Okke
  Schrijvers}.} \bibinfo{year}{2016}\natexlab{}.
\newblock \showarticletitle{Robust random cut forest based anomaly detection on
  streams}. In \bibinfo{booktitle}{{\em International conference on machine
  learning}}. PMLR, \bibinfo{pages}{2712--2721}.
\newblock


\bibitem[\protect\citeauthoryear{Han}{Han}{2020}]%
        {tm_github}
\bibfield{author}{\bibinfo{person}{Dongqi Han}.}
  \bibinfo{year}{2020}\natexlab{}.
\newblock \bibinfo{title}{Traffic Manipulator}.
\newblock
  \bibinfo{howpublished}{\url{https://github.com/dongtsi/TrafficManipulator/tree/99807048c61dea548c55bc2720ea31369bc90fd1}}.
    (\bibinfo{year}{2020}).
\newblock
\newblock
\shownote{Last Accessed: 2021-11-10.}


\bibitem[\protect\citeauthoryear{Han, Wang, Zhong, Chen, Yang, Lu, Shi, and
  Yin}{Han et~al\mbox{.}}{2020}]%
        {han2020practical}
\bibfield{author}{\bibinfo{person}{Dongqi Han}, \bibinfo{person}{Zhiliang
  Wang}, \bibinfo{person}{Ying Zhong}, \bibinfo{person}{Wenqi Chen},
  \bibinfo{person}{Jiahai Yang}, \bibinfo{person}{Shuqiang Lu},
  \bibinfo{person}{Xingang Shi}, {and} \bibinfo{person}{Xia Yin}.}
  \bibinfo{year}{2020}\natexlab{}.
\newblock \showarticletitle{Practical Traffic-space Adversarial Attacks on
  Learning-based NIDSs}.
\newblock \bibinfo{journal}{{\em arXiv preprint arXiv:2005.07519\/}}
  (\bibinfo{year}{2020}).
\newblock


\bibitem[\protect\citeauthoryear{Hashemi, Cusack, and Keller}{Hashemi
  et~al\mbox{.}}{2019}]%
        {hashemi2019towards}
\bibfield{author}{\bibinfo{person}{Mohammad~J Hashemi}, \bibinfo{person}{Greg
  Cusack}, {and} \bibinfo{person}{Eric Keller}.}
  \bibinfo{year}{2019}\natexlab{}.
\newblock \showarticletitle{Towards Evaluation of NIDSs in Adversarial
  Setting}. In \bibinfo{booktitle}{{\em Proceedings of the 3rd ACM CoNEXT
  Workshop on Big DAta, Machine Learning and Artificial Intelligence for Data
  Communication Networks}}. \bibinfo{pages}{14--21}.
\newblock


\bibitem[\protect\citeauthoryear{He, Zhang, Ren, and Sun}{He
  et~al\mbox{.}}{2016}]%
        {he2016identity}
\bibfield{author}{\bibinfo{person}{Kaiming He}, \bibinfo{person}{Xiangyu
  Zhang}, \bibinfo{person}{Shaoqing Ren}, {and} \bibinfo{person}{Jian Sun}.}
  \bibinfo{year}{2016}\natexlab{}.
\newblock \showarticletitle{Identity mappings in deep residual networks}. In
  \bibinfo{booktitle}{{\em European conference on computer vision}}. Springer,
  \bibinfo{pages}{630--645}.
\newblock


\bibitem[\protect\citeauthoryear{Homoliak, Teknos, Ochoa, Breitenbacher,
  Hosseini, and Hanacek}{Homoliak et~al\mbox{.}}{2018}]%
        {homoliak2018improving}
\bibfield{author}{\bibinfo{person}{Ivan Homoliak}, \bibinfo{person}{Martin
  Teknos}, \bibinfo{person}{Mart{\'\i}n Ochoa}, \bibinfo{person}{Dominik
  Breitenbacher}, \bibinfo{person}{Saeid Hosseini}, {and} \bibinfo{person}{Petr
  Hanacek}.} \bibinfo{year}{2018}\natexlab{}.
\newblock \showarticletitle{Improving network intrusion detection classifiers
  by non-payload-based exploit-independent obfuscations: An adversarial
  approach}.
\newblock \bibinfo{journal}{{\em arXiv preprint arXiv:1805.02684\/}}
  (\bibinfo{year}{2018}).
\newblock


\bibitem[\protect\citeauthoryear{Hu and Tan}{Hu and Tan}{2017}]%
        {hu2017generating}
\bibfield{author}{\bibinfo{person}{Weiwei Hu} {and} \bibinfo{person}{Ying
  Tan}.} \bibinfo{year}{2017}\natexlab{}.
\newblock \showarticletitle{Generating adversarial malware examples for
  black-box attacks based on gan}.
\newblock \bibinfo{journal}{{\em arXiv preprint arXiv:1702.05983\/}}
  (\bibinfo{year}{2017}).
\newblock


\bibitem[\protect\citeauthoryear{Ibitoye, Shafiq, and Matrawy}{Ibitoye
  et~al\mbox{.}}{2019}]%
        {ibitoye2019analyzing}
\bibfield{author}{\bibinfo{person}{Olakunle Ibitoye}, \bibinfo{person}{Omair
  Shafiq}, {and} \bibinfo{person}{Ashraf Matrawy}.}
  \bibinfo{year}{2019}\natexlab{}.
\newblock \showarticletitle{Analyzing adversarial attacks against deep learning
  for intrusion detection in IoT networks}. In \bibinfo{booktitle}{{\em 2019
  IEEE Global Communications Conference (GLOBECOM)}}. IEEE,
  \bibinfo{pages}{1--6}.
\newblock


\bibitem[\protect\citeauthoryear{Kennedy and Eberhart}{Kennedy and
  Eberhart}{1995}]%
        {kennedy1995particle}
\bibfield{author}{\bibinfo{person}{James Kennedy} {and}
  \bibinfo{person}{Russell Eberhart}.} \bibinfo{year}{1995}\natexlab{}.
\newblock \showarticletitle{Particle swarm optimization}. In
  \bibinfo{booktitle}{{\em Proceedings of ICNN'95-International Conference on
  Neural Networks}}, Vol.~\bibinfo{volume}{4}. IEEE,
  \bibinfo{pages}{1942--1948}.
\newblock


\bibitem[\protect\citeauthoryear{Kohonen and Oja}{Kohonen and Oja}{1998}]%
        {kohonen1998visual}
\bibfield{author}{\bibinfo{person}{Teuvo Kohonen} {and} \bibinfo{person}{Erkki
  Oja}.} \bibinfo{year}{1998}\natexlab{}.
\newblock \showarticletitle{Visual feature analysis by the self-organising
  maps}.
\newblock \bibinfo{journal}{{\em Neural Computing \& Applications\/}}
  \bibinfo{volume}{7}, \bibinfo{number}{3} (\bibinfo{year}{1998}),
  \bibinfo{pages}{273--286}.
\newblock


\bibitem[\protect\citeauthoryear{Kuppa, Grzonkowski, Asghar, and
  Le{-}Khac}{Kuppa et~al\mbox{.}}{2019}]%
        {kuppa2019black}
\bibfield{author}{\bibinfo{person}{Aditya Kuppa}, \bibinfo{person}{Slawomir
  Grzonkowski}, \bibinfo{person}{Muhammad~Rizwan Asghar}, {and}
  \bibinfo{person}{Nhien{-}An Le{-}Khac}.} \bibinfo{year}{2019}\natexlab{}.
\newblock \showarticletitle{Black Box Attacks on Deep Anomaly Detectors}. In
  \bibinfo{booktitle}{{\em Proceedings of the 14th International Conference on
  Availability, Reliability and Security, {ARES} 2019, Canterbury, UK, August
  26-29, 2019}}. \bibinfo{publisher}{{ACM}}, \bibinfo{address}{Canterbury, UK},
  \bibinfo{pages}{21:1--21:10}.
\newblock
\showDOI{%
\url{https://doi.org/10.1145/3339252.3339266}}


\bibitem[\protect\citeauthoryear{Kurakin, Goodfellow, and Bengio}{Kurakin
  et~al\mbox{.}}{2016}]%
        {kurakin2016adversarial}
\bibfield{author}{\bibinfo{person}{Alexey Kurakin}, \bibinfo{person}{Ian
  Goodfellow}, {and} \bibinfo{person}{Samy Bengio}.}
  \bibinfo{year}{2016}\natexlab{}.
\newblock \showarticletitle{Adversarial examples in the physical world}.
\newblock \bibinfo{journal}{{\em arXiv preprint arXiv:1607.02533\/}}
  (\bibinfo{year}{2016}).
\newblock


\bibitem[\protect\citeauthoryear{LeCun, Bengio, and Hinton}{LeCun
  et~al\mbox{.}}{2015}]%
        {lecun2015deep}
\bibfield{author}{\bibinfo{person}{Yann LeCun}, \bibinfo{person}{Yoshua
  Bengio}, {and} \bibinfo{person}{Geoffrey Hinton}.}
  \bibinfo{year}{2015}\natexlab{}.
\newblock \showarticletitle{Deep learning}.
\newblock \bibinfo{journal}{{\em nature\/}} \bibinfo{volume}{521},
  \bibinfo{number}{7553} (\bibinfo{year}{2015}), \bibinfo{pages}{436--444}.
\newblock


\bibitem[\protect\citeauthoryear{Lin, Shi, and Xue}{Lin et~al\mbox{.}}{2018}]%
        {lin2018idsgan}
\bibfield{author}{\bibinfo{person}{Zilong Lin}, \bibinfo{person}{Yong Shi},
  {and} \bibinfo{person}{Zhi Xue}.} \bibinfo{year}{2018}\natexlab{}.
\newblock \showarticletitle{Idsgan: Generative adversarial networks for attack
  generation against intrusion detection}.
\newblock \bibinfo{journal}{{\em arXiv preprint arXiv:1809.02077\/}}
  (\bibinfo{year}{2018}).
\newblock


\bibitem[\protect\citeauthoryear{Madry, Makelov, Schmidt, Tsipras, and
  Vladu}{Madry et~al\mbox{.}}{2017}]%
        {madry2017towards}
\bibfield{author}{\bibinfo{person}{Aleksander Madry},
  \bibinfo{person}{Aleksandar Makelov}, \bibinfo{person}{Ludwig Schmidt},
  \bibinfo{person}{Dimitris Tsipras}, {and} \bibinfo{person}{Adrian Vladu}.}
  \bibinfo{year}{2017}\natexlab{}.
\newblock \showarticletitle{Towards deep learning models resistant to
  adversarial attacks}.
\newblock \bibinfo{journal}{{\em arXiv preprint arXiv:1706.06083\/}}
  (\bibinfo{year}{2017}).
\newblock


\bibitem[\protect\citeauthoryear{Meng and Chen}{Meng and Chen}{2017}]%
        {meng2017magnet}
\bibfield{author}{\bibinfo{person}{Dongyu Meng} {and} \bibinfo{person}{Hao
  Chen}.} \bibinfo{year}{2017}\natexlab{}.
\newblock \showarticletitle{Magnet: a two-pronged defense against adversarial
  examples}. In \bibinfo{booktitle}{{\em Proceedings of the 2017 ACM SIGSAC
  conference on computer and communications security}}.
  \bibinfo{pages}{135--147}.
\newblock


\bibitem[\protect\citeauthoryear{Mirjalili, Mirjalili, and Lewis}{Mirjalili
  et~al\mbox{.}}{2014}]%
        {mirjalili2014grey}
\bibfield{author}{\bibinfo{person}{Seyedali Mirjalili},
  \bibinfo{person}{Seyed~Mohammad Mirjalili}, {and} \bibinfo{person}{Andrew
  Lewis}.} \bibinfo{year}{2014}\natexlab{}.
\newblock \showarticletitle{Grey wolf optimizer}.
\newblock \bibinfo{journal}{{\em Advances in engineering software\/}}
  \bibinfo{volume}{69} (\bibinfo{year}{2014}), \bibinfo{pages}{46--61}.
\newblock


\bibitem[\protect\citeauthoryear{Mirsky}{Mirsky}{2020}]%
        {kitsune_git}
\bibfield{author}{\bibinfo{person}{Yisroel Mirsky}.}
  \bibinfo{year}{2020}\natexlab{}.
\newblock \bibinfo{title}{Kitsune-py}.
\newblock \bibinfo{howpublished}{\url{https://github.com/ymirsky/Kitsune-py}}.
   (\bibinfo{year}{2020}).
\newblock
\newblock
\shownote{Accessed: 2020-9-15.}


\bibitem[\protect\citeauthoryear{Mirsky, Doitshman, Elovici, and
  Shabtai}{Mirsky et~al\mbox{.}}{2018}]%
        {mirsky2018kitsune}
\bibfield{author}{\bibinfo{person}{Yisroel Mirsky}, \bibinfo{person}{Tomer
  Doitshman}, \bibinfo{person}{Yuval Elovici}, {and} \bibinfo{person}{Asaf
  Shabtai}.} \bibinfo{year}{2018}\natexlab{}.
\newblock \showarticletitle{Kitsune: an ensemble of autoencoders for online
  network intrusion detection}.
\newblock \bibinfo{journal}{{\em arXiv preprint arXiv:1802.09089\/}}
  (\bibinfo{year}{2018}).
\newblock


\bibitem[\protect\citeauthoryear{Moosavi-Dezfooli, Fawzi, and
  Frossard}{Moosavi-Dezfooli et~al\mbox{.}}{2016}]%
        {moosavi2016deepfool}
\bibfield{author}{\bibinfo{person}{Seyed-Mohsen Moosavi-Dezfooli},
  \bibinfo{person}{Alhussein Fawzi}, {and} \bibinfo{person}{Pascal Frossard}.}
  \bibinfo{year}{2016}\natexlab{}.
\newblock \showarticletitle{Deepfool: a simple and accurate method to fool deep
  neural networks}. In \bibinfo{booktitle}{{\em Proceedings of the IEEE
  conference on computer vision and pattern recognition}}.
  \bibinfo{pages}{2574--2582}.
\newblock


\bibitem[\protect\citeauthoryear{Pang, Xu, Du, Chen, and Zhu}{Pang
  et~al\mbox{.}}{2019}]%
        {pang2019improving}
\bibfield{author}{\bibinfo{person}{Tianyu Pang}, \bibinfo{person}{Kun Xu},
  \bibinfo{person}{Chao Du}, \bibinfo{person}{Ning Chen}, {and}
  \bibinfo{person}{Jun Zhu}.} \bibinfo{year}{2019}\natexlab{}.
\newblock \showarticletitle{Improving adversarial robustness via promoting
  ensemble diversity}. In \bibinfo{booktitle}{{\em International Conference on
  Machine Learning}}. PMLR, \bibinfo{pages}{4970--4979}.
\newblock


\bibitem[\protect\citeauthoryear{Papernot, McDaniel, and Goodfellow}{Papernot
  et~al\mbox{.}}{2016a}]%
        {papernot2016transferability}
\bibfield{author}{\bibinfo{person}{Nicolas Papernot}, \bibinfo{person}{Patrick
  McDaniel}, {and} \bibinfo{person}{Ian Goodfellow}.}
  \bibinfo{year}{2016}\natexlab{a}.
\newblock \showarticletitle{Transferability in machine learning: from phenomena
  to black-box attacks using adversarial samples}.
\newblock \bibinfo{journal}{{\em arXiv preprint arXiv:1605.07277\/}}
  (\bibinfo{year}{2016}).
\newblock


\bibitem[\protect\citeauthoryear{Papernot, McDaniel, Jha, Fredrikson, Celik,
  and Swami}{Papernot et~al\mbox{.}}{2016b}]%
        {papernot2016limitations}
\bibfield{author}{\bibinfo{person}{Nicolas Papernot}, \bibinfo{person}{Patrick
  McDaniel}, \bibinfo{person}{Somesh Jha}, \bibinfo{person}{Matt Fredrikson},
  \bibinfo{person}{Z~Berkay Celik}, {and} \bibinfo{person}{Ananthram Swami}.}
  \bibinfo{year}{2016}\natexlab{b}.
\newblock \showarticletitle{The limitations of deep learning in adversarial
  settings}. In \bibinfo{booktitle}{{\em 2016 IEEE European symposium on
  security and privacy (EuroS\&P)}}. IEEE, \bibinfo{pages}{372--387}.
\newblock


\bibitem[\protect\citeauthoryear{Pierazzi, Pendlebury, Cortellazzi, and
  Cavallaro}{Pierazzi et~al\mbox{.}}{2020}]%
        {pierazzi2020intriguing}
\bibfield{author}{\bibinfo{person}{Fabio Pierazzi}, \bibinfo{person}{Feargus
  Pendlebury}, \bibinfo{person}{Jacopo Cortellazzi}, {and}
  \bibinfo{person}{Lorenzo Cavallaro}.} \bibinfo{year}{2020}\natexlab{}.
\newblock \showarticletitle{Intriguing properties of adversarial ml attacks in
  the problem space}. In \bibinfo{booktitle}{{\em 2020 IEEE Symposium on
  Security and Privacy (SP)}}. IEEE, \bibinfo{pages}{1332--1349}.
\newblock


\bibitem[\protect\citeauthoryear{Piplai, Chukkapalli, and Joshi}{Piplai
  et~al\mbox{.}}{2020}]%
        {piplai2020nattack}
\bibfield{author}{\bibinfo{person}{Aritran Piplai}, \bibinfo{person}{Sai
  Sree~Laya Chukkapalli}, {and} \bibinfo{person}{Anupam Joshi}.}
  \bibinfo{year}{2020}\natexlab{}.
\newblock \showarticletitle{NAttack! Adversarial Attacks to bypass a GAN based
  classifier trained to detect Network intrusion}.
\newblock \bibinfo{journal}{{\em arXiv preprint arXiv:2002.08527\/}}
  (\bibinfo{year}{2020}).
\newblock


\bibitem[\protect\citeauthoryear{Sak, Senior, and Beaufays}{Sak
  et~al\mbox{.}}{2014}]%
        {sak2014long}
\bibfield{author}{\bibinfo{person}{Ha{\c{s}}im Sak}, \bibinfo{person}{Andrew
  Senior}, {and} \bibinfo{person}{Fran{\c{c}}oise Beaufays}.}
  \bibinfo{year}{2014}\natexlab{}.
\newblock \showarticletitle{Long short-term memory based recurrent neural
  network architectures for large vocabulary speech recognition}.
\newblock \bibinfo{journal}{{\em arXiv preprint arXiv:1402.1128\/}}
  (\bibinfo{year}{2014}).
\newblock


\bibitem[\protect\citeauthoryear{Sengupta, Chakraborti, and
  Kambhampati}{Sengupta et~al\mbox{.}}{2019}]%
        {sengupta2019mtdeep}
\bibfield{author}{\bibinfo{person}{Sailik Sengupta}, \bibinfo{person}{Tathagata
  Chakraborti}, {and} \bibinfo{person}{Subbarao Kambhampati}.}
  \bibinfo{year}{2019}\natexlab{}.
\newblock \showarticletitle{{MTDeep}: boosting the security of deep neural nets
  against adversarial attacks with moving target defense}. In
  \bibinfo{booktitle}{{\em International Conference on Decision and Game Theory
  for Security}}. Springer, \bibinfo{pages}{479--491}.
\newblock


\bibitem[\protect\citeauthoryear{Sharafaldin, Lashkari, and
  Ghorbani}{Sharafaldin et~al\mbox{.}}{2018}]%
        {sharafaldin2018toward}
\bibfield{author}{\bibinfo{person}{Iman Sharafaldin},
  \bibinfo{person}{Arash~Habibi Lashkari}, {and} \bibinfo{person}{Ali~A
  Ghorbani}.} \bibinfo{year}{2018}\natexlab{}.
\newblock \showarticletitle{Toward generating a new intrusion detection dataset
  and intrusion traffic characterization.}. In \bibinfo{booktitle}{{\em
  ICISSP}}. \bibinfo{pages}{108--116}.
\newblock


\bibitem[\protect\citeauthoryear{Sommer and Paxson}{Sommer and Paxson}{2010}]%
        {sommer2010outside}
\bibfield{author}{\bibinfo{person}{Robin Sommer} {and} \bibinfo{person}{Vern
  Paxson}.} \bibinfo{year}{2010}\natexlab{}.
\newblock \showarticletitle{Outside the closed world: On using machine learning
  for network intrusion detection}. In \bibinfo{booktitle}{{\em 2010 IEEE
  Symposium on Security and Privacy (SP)}}. IEEE, \bibinfo{pages}{305--316}.
\newblock


\bibitem[\protect\citeauthoryear{Storn and Price}{Storn and Price}{1997}]%
        {storn1997differential}
\bibfield{author}{\bibinfo{person}{Rainer Storn} {and} \bibinfo{person}{Kenneth
  Price}.} \bibinfo{year}{1997}\natexlab{}.
\newblock \showarticletitle{Differential evolution--a simple and efficient
  heuristic for global optimization over continuous spaces}.
\newblock \bibinfo{journal}{{\em Journal of global optimization\/}}
  \bibinfo{volume}{11}, \bibinfo{number}{4} (\bibinfo{year}{1997}),
  \bibinfo{pages}{341--359}.
\newblock


\bibitem[\protect\citeauthoryear{Szegedy, Zaremba, Sutskever, Bruna, Erhan,
  Goodfellow, and Fergus}{Szegedy et~al\mbox{.}}{2013}]%
        {szegedy2013intriguing}
\bibfield{author}{\bibinfo{person}{Christian Szegedy},
  \bibinfo{person}{Wojciech Zaremba}, \bibinfo{person}{Ilya Sutskever},
  \bibinfo{person}{Joan Bruna}, \bibinfo{person}{Dumitru Erhan},
  \bibinfo{person}{Ian Goodfellow}, {and} \bibinfo{person}{Rob Fergus}.}
  \bibinfo{year}{2013}\natexlab{}.
\newblock \showarticletitle{Intriguing properties of neural networks}.
\newblock \bibinfo{journal}{{\em arXiv preprint arXiv:1312.6199\/}}
  (\bibinfo{year}{2013}).
\newblock


\bibitem[\protect\citeauthoryear{Trevillie}{Trevillie}{2018}]%
        {magnet_github}
\bibfield{author}{\bibinfo{person}{Trevillie}.}
  \bibinfo{year}{2018}\natexlab{}.
\newblock \bibinfo{title}{MagNet}.
\newblock \bibinfo{howpublished}{\url{https://github.com/Trevillie/MagNet}}.
  (\bibinfo{year}{2018}).
\newblock
\newblock
\shownote{Last Accessed: 2021-9-23.}


\bibitem[\protect\citeauthoryear{Vettigli}{Vettigli}{2018}]%
        {vettigliminisom}
\bibfield{author}{\bibinfo{person}{Giuseppe Vettigli}.}
  \bibinfo{year}{2018}\natexlab{}.
\newblock \bibinfo{title}{MiniSom: minimalistic and NumPy-based implementation
  of the Self Organizing Map}.
\newblock   (\bibinfo{year}{2018}).
\newblock
\showURL{%
\url{https://github.com/JustGlowing/minisom/}}


\bibitem[\protect\citeauthoryear{Xu, Evans, and Qi}{Xu et~al\mbox{.}}{2017}]%
        {xu2017feature}
\bibfield{author}{\bibinfo{person}{Weilin Xu}, \bibinfo{person}{David Evans},
  {and} \bibinfo{person}{Yanjun Qi}.} \bibinfo{year}{2017}\natexlab{}.
\newblock \showarticletitle{Feature squeezing: Detecting adversarial examples
  in deep neural networks}.
\newblock \bibinfo{journal}{{\em arXiv preprint arXiv:1704.01155\/}}
  (\bibinfo{year}{2017}).
\newblock


\end{thebibliography}

\appendix
\renewcommand{\thefigure}{A\arabic{figure}}
\setcounter{figure}{0}
\setcounter{table}{0}
\renewcommand{\thetable}{A\arabic{table}}

\newpage
\section{Experiment Hyperparameters}
\label{sec:hyperparameters}
We have used parameters shown in Table \ref{tab:constant} in our experiments. 
\begin{table}[h]
\small
\caption{\textit{Liuer Mihou's} hyperparameters used in all of the experiments.}
\label{tab:constant}
\begin{tabularx}{\linewidth}{lcX}
\toprule
\textbf{Variable} & \textbf{Value} & \textbf{Description}                                                \\ \midrule
n\_particles      & 20    & Population size of search algorithm                        \\
iterations        & 30    & Maximum number of evolutions of search algorithm           \\
mutation\_factor  & 0.8   & $\alpha$ used in differential evolution                    \\
cross\_p          & 0.7   & $p_r$ used in differential evolution                       \\
mutate\_prob      & 0.5   & Probability to apply differential evolution instead of PSO \\
max\_time\_window & 1 s     & Upper bound for packet delay       \\
max\_packet\_size & 1514 B  & Upper bound for packet size                          \\
max\_craft\_pkt   & 5     & Upper bound for packet injection                    \\
mutation probability   & 0.5     & Probability of updating with PSO              \\\bottomrule
\end{tabularx}
\end{table}

\section{Auxiliary Experiments}
\label{sec:perf_results}
In this section, we provide the results of our auxiliary experiments. 
Our auxiliary experiments include optimisation with PSO, finding optimal combinations of search algorithm and payload assignment, experiments with Traffic Manipulator \cite{han2020practical}, and performance of common outlier detection algorithms in NIDS.

\begin{figure*}[ht]
    \centering
    \begin{subfigure}{0.3\linewidth}
    \centering
        \includegraphics[width=\linewidth]{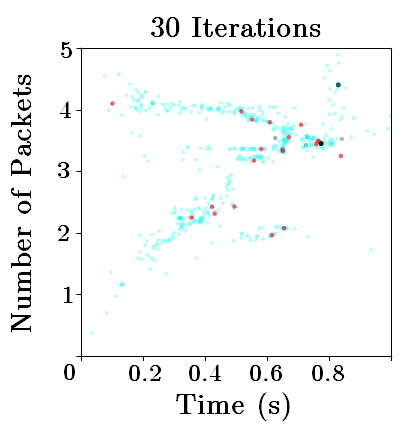}
    \caption{RA}
    \label{fig:2d_trace}
    \end{subfigure}
    \begin{subfigure}{0.3\linewidth}
    \centering
        \includegraphics[width=\linewidth]{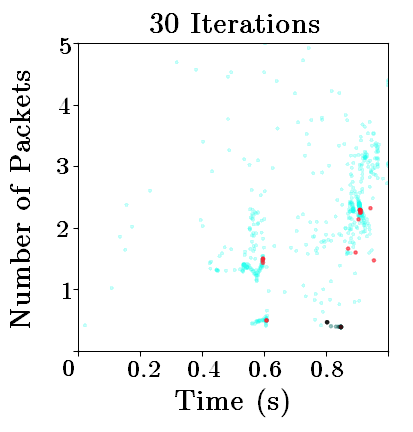}
    \caption{UA}
    \label{fig:3d_trace}
    \end{subfigure}
    \caption{Particle positions ($t_m, n_c$) during the search algorithm. 
    Light blue points are visited positions of the particles, red points are the position of the last iteration, and black points are locations of current and previous global best positions: 
    \textbf{(a)} RA with PSO search algorithm; 
    \textbf{(b)} UA with PSO search algorithm.}
    \label{fig:trace_comp}
\end{figure*}

\subsection{PSO Optimisation}
\label{sec:opt_with_pso}
We first conducted experiments to measure the performance of Liuer Mihou using vanilla PSO. 
We found that PSO causes the particles' position to stagnate and move only around its neighbourhood, reducing its exploration capabilities and finding locally optimal solutions. 
The stagnating particle problem is a well-known problem caused by the particle's current position being identical to its personal best position and neighbourhood best position. 
Under such circumstances, the cognitive and social terms are close to zero, and after few iterations, the inertia weight will tend to 0, resulting in a stagnating particle \cite{engelbrecht2013particle}.

To overcome this effect, we combine DE with PSO. 
In each iteration, the particle has 50\% chance of moving according to PSO and 50\% chance of moving according to DE. 
As a result, the social term of velocity changes stochastically in each iteration and reduces the probability of stagnating particles. 
Our experiments in the next section objectively compare the performance between pure PSO, pure DE and PSO-DE.

\subsection{Optimal Combination Search}
\label{sec:finding_opt_comb}

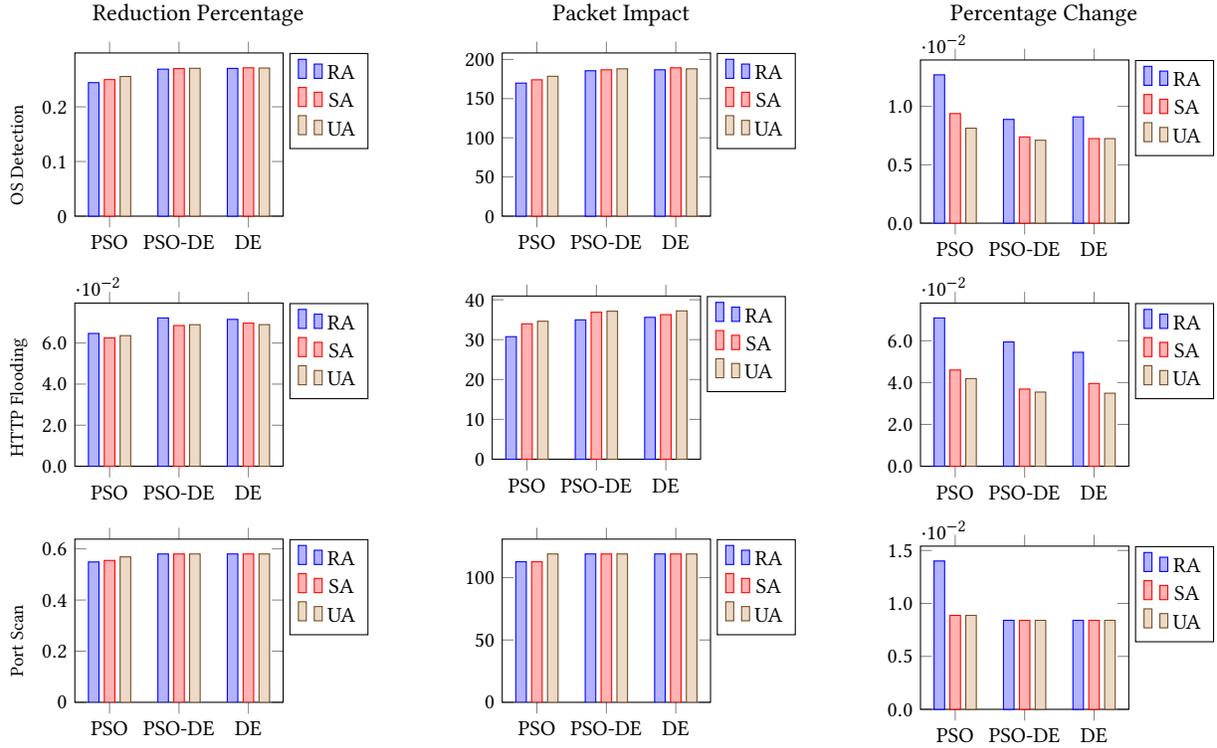
\begin{figure*}[t]
\centering
\setlength\tabcolsep{1pt}
\setkeys{Gin}{width=\hsize}
\settowidth\rotheadsize{HTTP Flooding}
\begin{tabularx}{0.9\linewidth}{lY@{\hskip 0.5in}Y@{\hskip 0.5in}Y}
& Reduction Percentage & Packet Impact & Percentage Change \\ \addlinespace[2pt]
\rothead{\centering OS Detection}  &   
\begin{tikzpicture}[thick,scale=1, every node/.style={scale=0.9}]
        \begin{axis}[
            ybar,
            ymin=0,
            enlarge x limits=0.25,
            bar width=4,
            width=\linewidth,
            symbolic x coords={PSO, PSO-DE, DE},
            legend pos=outer north east=outer north east,
            y tick label style={
        /pgf/number format/.cd,
        fixed,
        precision=1,
        /tikz/.cd
    },
            xtick=data
          ]
            \addplot coordinates {
                (PSO,  0.2442573899)
                (PSO-DE,  0.2689314164)
                (DE, 0.2703269873)
            };
            \addplot coordinates {
                (PSO,  0.2498169827)
                (PSO-DE,  0.270002507)
                (DE, 0.2713975823)
            };
            \addplot coordinates {
                (PSO, 0.2555195541)
                (PSO-DE, 0.2705139845)
                (DE, 0.2710006894)
            };
        \legend{RA, SA, UA}
        \end{axis}
    \end{tikzpicture}
                        &   
 \begin{tikzpicture}[thick,scale=1, every node/.style={scale=0.9}]
        \begin{axis}[
            ybar,
            ymin=0,
            enlarge x limits=0.25,
            width=\linewidth,
            bar width=4,
            symbolic x coords={PSO, PSO-DE, DE},
            legend pos=outer north east,
            xtick=data
          ]
            \addplot coordinates {
                (PSO, 169.8424242)
                (PSO-DE, 185.589404)
                (DE, 186.8266667)
            };
            \addplot coordinates {
               (PSO, 174.0621118)
                (PSO-DE, 186.8266667)
                (DE, 189.3513514)
            };
            \addplot coordinates {
               (PSO, 178.4968153)
                (PSO-DE, 188.0805369)
                (DE, 188.0805369)
            };
        \legend{RA, SA, UA}
        \end{axis}
    \end{tikzpicture}  
                        &   
    \begin{tikzpicture}[thick,scale=1, every node/.style={scale=0.9}]
        \begin{axis}[
            ybar,
            ymin=0,
            enlarge x limits=0.25,
            width=\linewidth,
            bar width=4,
            symbolic x coords={PSO, PSO-DE, DE},
            legend pos=outer north east,
                        y tick label style={
        /pgf/number format/.cd,
        fixed,
        fixed zerofill,
        precision=1,
        /tikz/.cd
    },
            xtick=data
          ]
            \addplot coordinates {
                (PSO, 0.01270339709)
                (PSO-DE, 0.008885241222)
                (DE, 0.00909934342)
            };
            \addplot coordinates {
               (PSO, 0.009384813017)
                (PSO-DE, 0.007386525835)
                (DE, 0.007243791036)
            };
            \addplot coordinates {
               (PSO, 0.008135883528)
                (PSO-DE, 0.007101056238)
                (DE, 0.007243791036)
            };
        \legend{RA, SA, UA}
        \end{axis}
    \end{tikzpicture}    \\  \addlinespace[2pt]
\rothead{\centering HTTP Flooding} &    
\begin{tikzpicture}[thick,scale=1, every node/.style={scale=0.9}]
        \begin{axis}[
            ybar,
            ymin=0,
            enlarge x limits=0.25,
            width=\linewidth,
            bar width=4,
            symbolic x coords={PSO, PSO-DE, DE},
            legend pos=outer north east,
            y tick label style={
        /pgf/number format/.cd,
        fixed,
        fixed zerofill,
        precision=1,
        /tikz/.cd
    },
            xtick=data
          ]
            \addplot coordinates {
                (PSO,  0.06467034389)
                (PSO-DE,  0.07223851886)
                (DE, 0.07152610914)
            };
            \addplot coordinates {
                (PSO,  0.06251662129)
                (PSO-DE,  0.06852530487)
                (DE, 0.06969481193)
            };
            \addplot coordinates {
                (PSO, 0.06360669249)
                (PSO-DE,0.068933443)
                (DE, 0.06897211536)
            };
        \legend{RA, SA, UA}
        \end{axis}
    \end{tikzpicture}
                        &  
    \begin{tikzpicture}[thick,scale=1, every node/.style={scale=0.9}]
        \begin{axis}[
            ybar,
            ymin=0,
            enlarge x limits=0.25,
            width=\linewidth,
            bar width=4,
            symbolic x coords={PSO, PSO-DE, DE},
            legend pos=outer north east,
            xtick=data
          ]
            \addplot coordinates {
                (PSO, 30.746)
                (PSO-DE, 34.96)
                (DE, 35.586)
            };
            \addplot coordinates {
               (PSO, 33.965)
                (PSO-DE, 36.869)
                (DE, 36.27)
            };
            \addplot coordinates {
               (PSO, 34.611)
                (PSO-DE, 37.124)
                (DE, 37.212)
            };
        \legend{RA, SA, UA}
        \end{axis}
    \end{tikzpicture}
                        & 
    \begin{tikzpicture}[thick,scale=1, every node/.style={scale=0.9}]
        \begin{axis}[
            ybar,
            ymin=0,
            enlarge x limits=0.25,
            width=\linewidth,
            bar width=4,
            symbolic x coords={PSO, PSO-DE, DE},
                        y tick label style={
        /pgf/number format/.cd,
        fixed,
        fixed zerofill,
        precision=1,
        /tikz/.cd
    },
            legend pos=outer north east,
            xtick=data
          ]
            \addplot coordinates {
                (PSO, 0.07090353217)
                (PSO-DE,0.05941075515)
                (DE, 0.05445961895)
            };
            \addplot coordinates {
               (PSO, 0.04598851759)
                (PSO-DE, 0.03688735794)
                (DE, 0.03956437827)
            };
            \addplot coordinates {
               (PSO, 0.04189419549)
                (PSO-DE, 0.0354487663)
                (DE, 0.03490809416)
            };
        \legend{RA, SA, UA}
        \end{axis}
    \end{tikzpicture}\\  \addlinespace[2pt]
\rothead{\centering Port Scan}     &  
\begin{tikzpicture}[thick,scale=1, every node/.style={scale=0.9}]
        \begin{axis}[
            ybar,
            ymin=0,
            enlarge x limits=0.25,
            width=\linewidth,
            bar width=4,
            symbolic x coords={PSO, PSO-DE, DE},
            legend pos=outer north east,
            xtick=data
          ]
            \addplot coordinates {
                (PSO, 0.5490021298)
                (PSO-DE,  0.5802436968)
                (DE, 0.5802436968)
            };
            \addplot coordinates {
                (PSO,  0.5549331809)
                (PSO-DE,  0.5802436968)
                (DE, 0.5802436968)
            };
            \addplot coordinates {
                (PSO, 0.5687177198)
                (PSO-DE, 0.5802436968)
                (DE, 0.5802436968)
            };
        \legend{RA, SA, UA}
        \end{axis}
    \end{tikzpicture}
                        &   
\begin{tikzpicture}[thick,scale=1, every node/.style={scale=0.9}]
        \begin{axis}[
            ybar,
            ymin=0,
            enlarge x limits=0.25,
            width=\linewidth,
            bar width=4,
            symbolic x coords={PSO, PSO-DE, DE},
            legend pos=outer north east,
            xtick=data
          ]
            \addplot coordinates {
                (PSO, 112.7368421)
                (PSO-DE, 119)
                (DE, 119)
            };
            \addplot coordinates {
               (PSO, 112.7368421)
                (PSO-DE, 119)
                (DE, 119)
            };
            \addplot coordinates {
               (PSO, 119)
                (PSO-DE, 119)
                (DE, 119)
            };
        \legend{RA, SA, UA}
        \end{axis}
    \end{tikzpicture}
                        &       
\begin{tikzpicture}[thick,scale=1, every node/.style={scale=0.9}]
        \begin{axis}[
            ybar,
            ymin=0,
            bar width=4,
            enlarge x limits=0.25,
            width=\linewidth,
            symbolic x coords={PSO, PSO-DE, DE},
            legend pos=outer north east,
                        y tick label style={
        /pgf/number format/.cd,
        fixed,
        fixed zerofill,
        precision=1,
        /tikz/.cd
    },
            xtick=data
          ]
            \addplot coordinates {
                (PSO, 0.01400560224)
                (PSO-DE,0.008403361345)
                (DE, 0.008403361345)
            };
            \addplot coordinates {
               (PSO, 0.008870214753)
                (PSO-DE, 0.008403361345)
                (DE, 0.008403361345)
            };
            \addplot coordinates {
               (PSO, 0.008870214753)
                (PSO-DE, 0.008403361345)
                (DE, 0.008403361345)
            };
        \legend{RA, SA, UA}
        \end{axis}
    \end{tikzpicture}\\  \addlinespace[2pt]
\end{tabularx}
    \caption{Summary of experimental results: 
    Rows represent the three attacks and columns represent the three model metrics.
    }
\label{fig:results}
\end{figure*}

We wish to know which combination of payload assignment (Random Assignment (RA), Seeded Assignment (SA), and Uniform Assignment (UA)) and search algorithm (DE, PSO, PSO-DE) yields the best solution to our optimisation problem (Equation \ref{opt_max}). 

We have used three metrics to objectively measure the performance of each combination of search algorithm and payload and arrival assignment:
\begin{description}

    \item [Reduction Percentage (RP)] RP measures the average amount of reduction in anomaly score for each malicious packet. 
    A higher reduction percentage means the search algorithm can find adversarial packets with a lower cost value.

    \item [Packet Impact (PI)] PI measures the effect of adversarial mutation on the malicious packet as well as subsequent attack packets. 
    The sequential relationships between Kitsune's features mean that lowering one packet's anomaly score will reduce the subsequent packets' anomaly scores until the anomaly scores eventually climb back up over the threshold. 
    The number of subsequent attack packets that an adversarial packet lowers below the threshold is defined as the impact of the packet. 
    The packet impact metric measures the average packet impact of all malicious packets for an attack file. 
    A high packet impact indicates fewer malicious packets modified overall. 

    \item [Percentage Change (PC)] PC measures the number of changes that the \textit{Liuer Mihou} attack has to modify to make all attack packets bypass detection. 
    We did not explicitly minimise the number of packets modified in the attack formulation, so this metric is mainly used as a tie-breaker when models have similar RP and PI.

\end{description}

Figure \ref{fig:results} shows the RP, PI, and PC for all three attacks. 
Across all the attacks, the assignment method does not significantly impact the PI and RP given a particular search algorithm. 
However, RA has a significantly higher PC compared to all other assignment methods. 
PSO has the worst RP and PI in terms of the search algorithms; while. PSO-DE and DE have similar performance. 
Overall, UA assignment with the PSO-DE algorithm has the best result because of its high RP, high PI, and low PC for all attacks.

An interesting finding from our results is that RA has similar RP and PI compared to other assignment methods but have significantly higher PC. 
To better understand this phenomenon, we have compared the history of the particles' position during the same optimisation problem between RA and UA with PSO in Figure \ref{fig:trace_comp}. 

With RA, the particles move near the local neighbourhood but never converge to a single point, leaving much of the search space unexplored. 
On the other hand, particles with UA have searched a larger area and converged to a point within its local neighbourhood. 
The stagnation of particles reveals a weakness of RA, which is the lack of total order between any pair of positions. 
To illustrate, consider an extremely simplified scenario involving two particles $\psi_1, \psi_2$ within the same neighbourhood that has cost values $c(\psi_1)=0.3$ and $c(\psi_2)=0.5$ at iteration $i$. 
Under such circumstances, the PSO algorithm will move $\psi_2$ towards $\psi_1$. 
In the next iteration, $i+1$, the cost values may be completely different due to non-deterministic nature of RA, say, $c(\psi_1)=0.5$ and $c(\psi_2)=0.2$. 
The PSO algorithm will now move $\psi_1$ towards $\psi_2$. 
This process may repeat several times, resulting in particles moving around in circles and never leaves their local neighbourhood. 

Despite the low exploration capability of RA, it still finds a ``good enough'' solution compared to other assignments. 
The reason is also due to the stochastic nature of RA. 
When the particles move within a local neighbourhood, they repeatedly sample the distribution to find the lowest cost value. 
Since multiple optimal solutions exist, the search algorithm will eventually find an optimal solution with a high number of crafted packets.  

\newpage
\subsection{Traffic Manipulator Experiments}
\label{sec:exp_han}
We have experimented with Traffic Manipulator (TM) downloaded the open source implementation of TM on GitHub \cite{tm_github}.

Following the instructions on the GitHub page but using our own test data, we have obtained the results with Port Scan (PS), OS \& Service Detection (OD) and HTTP Flooding (HF) attacks, shown in Table \ref{tab:exp_tm}. 

The results show that after TM's manipulation, the malicious packets are less evasive. 
We suspect this might be due to our poor choice of mimic features since we naively chose 1000 random benign features as the mimic set. 
The authors claim that using a GAN to generate mimic features yield better results. 
However, the code for GAN is not available on GitHub.

\begin{table}[ht]
\caption{Evasiveness of Traffic Manipulator (TM) on our dataset.}
\label{tab:exp_tm}
\begin{tabularx}{\linewidth}{@{}XXXX@{}}
\toprule
Attack & MDR & ADR & AER \\ \midrule
PS & 0.807 & 0.947 & -0.174 \\
OD & 0.389 & 0.925 & -1.376 \\
HF & 0.980 & 0.991 & -0.011 \\ \bottomrule
\end{tabularx}
\end{table}

\subsection{Outlier Detection Algorithms Evaluation}
\label{sec:pre_transfer}
We have conducted experiments to find suitable outlier detection algorithms for conducting transferability analysis in Section \ref{sec:performance_analysis}. 
Table \ref{tab:other_ml_metric} shows the performance metrics of different outlier detection algorithms. 
We have used the sklearn package to create the models and used the default parameters for all of the algorithms and, where applicable, set the upper bound on the fraction of training errors to be 0.001. 

\begin{table}[ht]
\small
\caption{Performance of the different NIDSes.}
\label{tab:other_ml_metric}
\begin{tabularx}{\linewidth}{@{}YYYY@{}}
\toprule
\textbf{Attack} & \textbf{Algorithm} &  \textbf{TNR}  & \textbf{MDR} \\ \midrule

PS & Kitsune  & 1.000 & 0.736 \\

PS & SOM  & 0.993  & 0.894 \\

PS & LOF  & 0.999  & 0.982 \\

PS & RRCF  & 0.992  & 0.929 \\

PS & OCSVM  & 0.999  & 0.986 \\

PS & IF  & 0.999  & 0.000 \\

PS & EE  & 1.000  & 0.000 \\
\rowcolor[HTML]{EFEFEF} 

OD & Kitsune  & 1.000  & 0.406 \\
\rowcolor[HTML]{EFEFEF} 

OD & SOM  & 0.993  & 0.656 \\
\rowcolor[HTML]{EFEFEF} 

OD & LOF  & 0.999  & 0.954 \\
\rowcolor[HTML]{EFEFEF} 

OD & RRCF  & 0.995  & 0.642 \\
\rowcolor[HTML]{EFEFEF} 

OD & OCSVM  & 0.999  & 0.835 \\
\rowcolor[HTML]{EFEFEF} 

OD & IF  & 0.999  & 0.000 \\
\rowcolor[HTML]{EFEFEF} 

OD & EE  & 1.000  & 0.000 \\

HF & Kitsune  & 1.000  & 0.999 \\

HF & SOM  & 0.993  & 0.999 \\

HF & LOF  & 0.999  & 1.000 \\

HF & RRCF  & 0.978  & 1.000 \\

HF & OCSVM  & 0.999  & 1.000 \\

HF & IF  & 0.999  & 0.000 \\

HF & EE  & 1.000  & 0.000\\ \bottomrule

\end{tabularx}
\end{table}

Results show IF and EE have a high TNR of close to 1 but a low MDR rate of 0 for all three attacks, which suggests IF and EE classify all traffic as benign and not suitable for our dataset.

\newpage
\section{Supplementary Material}

\subsection{Number of Packets}

\begin{table}[h]
\small
\caption{The total number of malicious ($T_{mal}$), adversarial ($T_{adv}$) and replay ($T_{rep}$) packets for each attack.}
\label{tab:packet_num}
\begin{tabularx}{\linewidth}{@{}XXXX@{}}
\toprule
\textbf{Attack} & \multicolumn{1}{l}{\textbf{$T_{mal}$}} & \multicolumn{1}{l}{\textbf{$T_{adv}$}} & \multicolumn{1}{l}{\textbf{$T_{rep}$}} \\ \midrule
PS              & 2142                                   & 2142                                   & 2252                                   \\
OD              & 28024                                  & 28074                                  & 33571                                  \\
HF              & 640767                                 & 37440                                  & 36653                                  \\ \bottomrule
\end{tabularx}
\end{table}

\subsection{Symbols and Equations used in Metrics}
\label{sec:metrics_symbols}
\begin{table}[h]
\centering
\small
\caption{Summary of metrics used in experiments.}
\label{tab:metrics}
\begin{tabularx}{\linewidth}{lYY}
\toprule
\textbf{Metric Name} & \textbf{Formula} & \textbf{Type} \\ \midrule
True Negative Rate (TNR)  & $\displaystyle \frac{T_{\textrm{ben}}-\rho_{\textrm{ben}}}{T_{\textrm{ben}}} $  & Performance\\[1em]

Malcious Detection Rate (MDR) & $\displaystyle \rho_{\textrm{mal}}/T_{\textrm{mal}} $     & Performance  \\[1em]

Adversarial Detection Rate (ADR)          & $\displaystyle \rho_{\textrm{adv}}/T_{\textrm{adv}} $ & Evasion    \\[1em]

Replay Detection Rate (RDR)                 & $\displaystyle \rho_{\textrm{rep}}/T_{\textrm{rep}} $        & Evasion     \\[1em]

Adversarial Evasion Rate (AER)          & $ \displaystyle \frac{\displaystyle \textrm{MDR}-\textrm{ADR}}{ \displaystyle \textrm{MDR}} $ & Evasion    \\[1em]

Replay Evasion Rate (RER)                 & $\displaystyle \frac{\displaystyle \textrm{MDR}-\textrm{RDR}}{ \displaystyle\textrm{MDR}} $        & Evasion     \\[1em]
Relative Round-trip Delay (RRD)                 & $\displaystyle R_a/R_o $ & Semantic    \\[1em]
Relative Ports Scanned (RPS)              & $\displaystyle PS_a/PS_o $ & Semantic \\[1em]
Relative Time Delay (RTD) &                 $ \displaystyle t_a/t_o$ & Semantic \\ \bottomrule
\end{tabularx}
\end{table}

\begin{table}[h]
\small
\caption{Summary of symbols used in metrics.}
\label{tab:symbols}
\begin{tabularx}{\linewidth}{ll}
\toprule
\textbf{Symbol} & \textbf{Meaning} \\ \midrule

$\rho_{\textrm{ben}}$   & \# of positive packets detected in benign packets  \\

$T_{\textrm{ben}}$ & Total number of benign packets                   \\

$\rho_{\textrm{mal}}$   &  \# of positive packets detected in malicious packets \\

$T_{\textrm{mal}}$ & Total number of malicious packets              \\

$\rho_{\textrm{adv}}$   & \# of positive packets detected in adversarial packets\\

$T_{\textrm{adv}}$ & Total number of adversarial packets              \\

$\rho_{\textrm{rep}}$   & \# of positive packets detected in replay packets  \\

$T_{\textrm{rep}}$ & Total number of replayed packets              \\

$R_o, R_a$    & Response time of original and adversarial attack \\

$PS_o, PS_a$    & Ports detected by original and adversarial attacks\\

$t_o, t_a$    & Time taken for original and adversarial attacks\\ \bottomrule
\end{tabularx}
\end{table}


\end{document}